\documentclass[showpacs,showkeys,preprintnumbers]{revtex4}%
\usepackage{amsfonts}
\usepackage{amsmath}
\usepackage{graphicx}
\usepackage{dcolumn}
\usepackage{bm}
\usepackage{amssymb}%
\usepackage{setspace}
\usepackage{color,hyperref}
\providecommand{\U}[1]{\protect\rule{.1in}{.1in}}

\begin{document}

\title{$X(3872)$ as virtual companion pole of the charm-anticharm state $\chi
_{c1}(2P)$}
\author{Francesco Giacosa$^{\text{(a,b)}},$ Milena Piotrowska$^{\text{(a)}}$, Susana
Coito$^{\text{(a)}}$}
\affiliation{$^{\text{(a)}}$ Institute of Physics, Jan Kochanowski University, ul.
Swietokrzyska 15, 25-406 Kielce, Poland,}
\affiliation{$^{\text{(b)}}$Institute for Theoretical Physics, Goethe University,
Max-von-Laue-Str.\ 1, 60438 Frankfurt am Main, Germany.}

\begin{abstract}
We study the spectral function of the axial-vector charmonium state $\chi_{c1}(2P)$ coupled to $DD^{\ast}$ mesons, by employing a quantum field
theoretical approach: a pronounced enhancement close to the $D^{0}%
D^{\ast0}$ threshold, to be identified with the $X(3872)$, emerges. In the complex plane we find two poles: a pole for the broad seed state $\chi_{c1}(2P)$, and -in the easiest scenario- a virtual pole for the $X(3872)$. Thus, our approach describes both the seed state and the
dynamically generated $X(3872)$ simultaneously. In particular, it explains the most prominent, both molecular-like and quarkonium-like,  features of the $X(3872)$: its very small width (the decay into $D^{0}D^{\ast0}$ is predicted to be about 0.5 MeV), the enhanced radiative decay into $\psi(2S)\gamma$ w.r.t.
$\psi(1S)\gamma$, and the isospin breaking decay into $J/\psi\rho$ (thanks to $DD^{\ast}$ loops mediating this decay channel). At the same time, we aim to determine the pole position and the properties
of the charmonium seed state: quite interestingly, even if a pole is always
present, it is possible that there is no peak  corresponding to this state in
the spectral function, thus potentially explaining why the corresponding resonance could not
yet be seen in experiments.
\end{abstract}

\pacs{11.10.Ef, 12.40.Yx, 14.40.Gx}
\keywords{axial-vector charmonia, $X(3872)$, unquenching, effective QCD approaches}
\maketitle

\section{Introduction}

The axial-vector resonance $X(3872)$, discovered in 2003 by the Belle
collaboration \cite{belle} and later on confirmed by various experimental
collaborations \cite{pdg}, is the first one of a series so-called $X,Y,Z$ states, which do not fit into the simple quark-antiquark
picture (see the review papers
\cite{rev2016,brambilla,pillonireview,nielsen,langereview,olsenrev} and refs.~therein). The state $X(3872)$ is reported in the PDG \cite{pdg} under the name
$\chi_{c1}(3872)$: the average mass is $m_{X(3872)}^{PDG}=3871.69\pm0.17$ MeV,
while only an upper limit for the width is given: $\Gamma_{X(3872)}^{PDG}<1.2$
MeV (90\% CL).

The $X(3872)$ is quite unique, since it is very narrow and its mass is very
close to the $D^{0}D^{\ast0}$ threshold (it is not yet clear if slightly above
or below it). In the $J^{PC}=1^{++}$ sector, the quark model, using the standard Cornell potential, predicts the
existence of a charm-anticharm state $\chi_{c1}(2P)$ at about $3.95$
GeV \cite{isgur,eichten,ebert} (in spectroscopic notation: $n$ $^{2S+1}L_J=$  $2$ $^{3}P_{1}$),
which is too high to be straightforwardly identified with the $X(3872)$.

Some features of the $X(3872)$ are very well described by interpreting
it as a $D^{0}D^{\ast0}$ molecular state
\cite{kalashnikova,gutschedong,gamermann,stapleton,hanhart,kalashnikova2}: its mass is very
close to the $D^{0}D^{\ast0}$ threshold and the isospin-breaking decay $X(3872)\rightarrow
J/\psi\rho$ can be understood. (Note, we use $D^0D^{*0}$ as a shortcut for $D^0\bar{D}^{*0}+h.c.$ ). However, some other features, such as the radiative
decays and the prompt production of the $X(3872)$ in heavy ion collisions, are better described by a charm-anticharm (or another compact) structure
\cite{barnesgodfrey,achasovcharmonium,eichtenx3872}.  (In what concerns prompt
production, there is an ongoing debate, see Refs.~\cite{bignaminiprl,pilloniconundrum,atlasx3872,hanhartcomment,maianinote}). In
order to account for these different phenomenological aspects, models in which
both $\bar{c}c$ and $D^{0}D^{\ast0}$ enter in the wave function of the
$X(3872)$ were proposed \cite{ortega,coito3872,santopinto,ferretti,cardoso,takeuchi,xiao1,xiao2}.
In other approaches, the $X(3872)$ is described as a compact diquark-antidiquark
state, e.g.~Ref.~\cite{maianix3872, dubnicka, matheus}.

In this work, in line with previous works on the so-called companion poles (see
Refs.~\cite{boglione1,boglione2,tornqvist,dullemond,a0,soltysiak} for the
light sector and Refs.~\cite{3770,milenapeter} for the charmonium sector), we
follow a different and quite simple idea: instead of working with quark
degrees of freedom, a Lagrangian in which a single axial-vector field
$\chi_{c1}^{\mu}$, which we identify -in the non-interacting limit- as a $\bar{c}c$ seed state $\chi
_{c1}\equiv\chi_{c1}(2P)$ with a bare mass between $3.90$ and
$3.95$ GeV, is considered. This field couples strongly to the $D^{0}D^{\ast0}$ and $D^{+}D^{\ast-}$ meson pairs. As a consequence, the bare seed state is dressed by mesonic quantum fluctuations and  acquires a total decay width of about $80$ MeV, as predicted by the quark model. We
then study the dressed propagator and its imaginary part, which delivers its
spectral function. For a suitable but quite natural choice of the coupling of
$\chi_{c1}^{\mu}$ to $DD^{\ast}$, the spectral function of the charmonium
state $\chi_{c1}(2P)$ shows a strong enhancement close to the $D^{0}D^{\ast0}$
threshold. In the easiest scenario, that we put forward here, this enhancement
corresponds to a \textit{virtual} state, hence to a pole in the II Riemann
sheet (RS) on the real axis and below the lowest $D^{0}D^{\ast0}$ threshold.
This pole is interpreted as a companion pole. As expected, the original
`standard' seed pole in the III RS (above both thresholds) also
exists: its real part is between $3.95$ and $4$ GeV and its imaginary part is
about $35$ MeV, in agreement with the predictions of the quark model
\cite{isgur,barnesgodfrey}.

The scenario proposed in our work merges quite well both the charmonium and
the molecular behavior of the $X(3872)$. This result is achieved naturally by the mesonic
quantum fluctuations dressing the bare state and without considering explicitly the Fock
space with both $\bar{q}q$ and molecular components. It is important to stress
that there is only one spectral function, correctly normalized to $1$, hence
strictly speaking there is only `one object'. However, the shape of this
spectral function is non-trivial and two (relevant) poles on the complex plane
are present. Quite interestingly, for certain sets of parameters, the spectral
function shows only one peak close to threshold and no peak corresponding to
the seed state, thus possibly explaining why the $\bar{c}c$ seed state could
not yet be experimentally measured. 

Moreover, our study can easily explain why the strong decay into $D^{0}D^{\ast0}$ is dominant (predictions for this decay width are evaluated to be about 0.5 MeV).  Moreover, one can understand why radiative decays are in agreement
with the charmonium assignment: since $X(3872)$ is part of the spectral
function of the whole state $\chi_{c1}(2P)$ (originally a $\bar{c}c$ seed
state dressed by $DD^{\ast}$ loops), the coupling constants are basically the
same (see later for details), hence the decay into $\psi(2S)\gamma$ is larger
than the decay into $\psi(1S)\gamma$. For the very same reason, the prompt production of
the $X(3872)$ in heavy ion collisions is quite natural. However, due to
the close threshold and dressing, various `molecular-like' properties also emerge:
the ratio $X(3872)\rightarrow J/\psi\omega\rightarrow J/\psi\pi^{+}\pi^{-}%
\pi^{0}$ over $X(3872)\rightarrow J/\psi\rho\rightarrow J/\psi\pi^{+}\pi^{-}$
can be correctly described by taking into account the small difference between the $D^{0}D^{\ast0}$ and $D^{+}D^{\ast-}$ loop functions.

In the end, it should be stressed that within our approach the very existence of the $X(3872)$ is not possible without 
the seed charmonium state. 
If one sends the mass of the latter to infinity and/or reduce the interaction strength to $DD^{\ast}$ mesons, the peak 
associated to $X(3872)$ disappears. This is a feature shared also by works based on quantum mechanics 
in which the $X(3872)$ was treated as a stationary state defined by a multi-component wave-function with a $\bar{c}c$ 
coupled to meson-meson channels \cite{coito3872,cardoso}. 
As a consequence of this discussion, it is important to stress that the
present paper does not focus only on the state  $X(3872)$, but on its relation
with and on the properties of the corresponding seed charmonium state
$\chi_{c_{1}}(2P)$. We aim to evaluate the pole of this charm-anticharm state,
as well as some partial decay widths. Moreover, we shall also investigate if
and under which condition a peak on the spectral function appears for this
state. In one interesting scenario, even if the pole for this state is
present, the peak does not appear, since it is `washed away' by the $DD^{\ast
}$ loops dressing it. This feature could explain why the seed $c\bar{c}$ state
was not yet seen in experiments (and is indeed one of the most important
outcomes of our study).

The paper is organized as follows: in Sec.~\ref{s2} we present the model and
in\ Sec.~\ref{s3} the results, divided into strong decays, radiative decays, and
isospin-breaking decays. Finally, in\ Sec.~\ref{s4} we present our conclusions. In
addition, in the Appendix \ref{ap} the results for different parameter choices are reported.

\section{\label{s2}The model and its consequences}

\textit{The Lagrangian}: 
We consider the following Lagrangian that couples a
seed state $\chi_{c1}$ with quantum numbers $1^{++}$ (and $n$ $^{2S+1}L_J=$  $2$ $^{3}P_{1}$), to
$D^{0}\bar{D}^{\ast0}+h.c.$ and $D^{+}D^{\ast-}+h.c.$ meson pairs:
\begin{equation}
\mathcal{L}_{\chi_{c1}(2P)DD^{\ast}}=g_{\chi_{c1}DD^{\ast}}\chi_{c1,\mu
}\left[  D^{\ast0,\mu}\bar{D}^{0}+D^{\ast+,\mu}D^{-}+h.c.\right]  \text{ ,}%
\end{equation}
where:

(i) $g_{\chi_{c1}DD^{\ast}}$ is the coupling constant with dimension [Energy]
(because of isospin symmetry, it is the same in the neutral and charged channels).

(ii) The theory is regularized via a form factor $F_{\Lambda}(k)$, which takes
effectively into account the finite dimensions of the mesons and their
interactions (microscopically, there is a nonlocal triangle diagram involving
two $c$ quarks and one light quark \cite{nl}). The tree-level decay widths as
function of the `running mass' $m$ of $\chi_{c1}$ read:%

\begin{align}
\Gamma_{\chi_{c1}(2P)\rightarrow D^{\ast0}\bar{D}^{0}+h.c.}(m)  &
=2\frac{k(m,m_{D^{\ast0}},m_{D^{0}})}{8\pi m^{2}}\frac{g_{\chi_{c1}DD^{\ast}%
}^{2}}{3}\left(  3+\frac{k^{2}(m,m_{D^{\ast0}},m_{D^{0}})}{m_{D^{\ast0}}^{2}%
}\right)  F_{\Lambda}(k)\text{ ,}\label{dw1}\\
\Gamma_{\chi_{c1}(2P)\rightarrow D^{\ast+}D^{-}+h.c.}(m)  &  =2\frac
{k(m,m_{D^{\ast+}},m_{D^{+}})}{8\pi m^{2}}\frac{g_{\chi_{c1}DD^{\ast}}^{2}}%
{3}\left(  3+\frac{k^{2}(m,m_{D^{\ast+}},m_{D^{+}})}{m_{D^{\ast+}}^{2}%
}\right)  F_{\Lambda}(k)\text{ ,} \label{dw2}%
\end{align}
where $k\equiv k(m,m_{1},m_{2})$ is the modulus of the three-momentum of an
emitted particle ($m$ is the mass of the decaying particle and
$m_{1}$ and $m_{2}$ the masses of the decay products). 

 Note, it is clear that out of the local Lagrangian of
Eq. (1) the Feynman rules deliver the decay width without the form factor.
Then, while often a local Lagrangian such as Eq. (1) is written down for
simplicity, in order to obtain Eqs. (2) and (3) -together with the form
factor- directly from the Lagrangian, one has to render it nonlocal, see Refs.
\cite{nl,nl2,lupo,covariant}  and refs. therein for the explicit and detailed
treatment of this issue. Namely, the standard Feynman rules in the case of a
nonlocal Lagrangian delivers the form factor as the Fourier transform of a
function depending on the space-time distance between the decay products.

(iii) We choose a Gaussian form factor,
\begin{equation}
F_{\Lambda}(k)=e^{-\frac{2k^{2}(m,m_{1},m_{2})}{\Lambda^{2}}}\text{
,}\label{gaussff}%
\end{equation}
since it emerges quite naturally in the $^{3}P_{0}$ model from the overlap of
the mesonic wave functions (e.g. Refs.
\cite{segovia,segovia2,segoviarev,amslerclose,swanson3p0,3p0recent}). The
precise form of $F_{\Lambda}(k)$ has a small influence on the results, as long
as it is a smooth function which guarantees convergence, see the detailed
discussion in Refs.~\cite{soltysiak,3770,milenapeter,lupo}. As shown in\ Ref.~\cite{covariant}, covariance is fulfilled even when using a form factor that
cuts the three-momentum $k$ (such as in\ Eq. (\ref{gaussff})), provided that
this form is used only in the reference frame in which the decaying particle
is at rest. In the Appendix \ref{ap} we will also use, for comparison and for
completeness, a different vertex.

(iv) The numerical value of $\Lambda$ is quite important. Typically, it ranges
from $0.4$ to $0.8$ GeV \cite{soltysiak,a0,nl,3p0recent}. Note, $\Lambda$
should not be regarded as a cutoff of a fundamental theory, but as a parameter
which is inversely proportional to the radius of mesons. We shall start with
$500$ MeV (typical for mesonic objects \cite{soltysiak,milenapeter}), but as
we find qualitatively similar results by varying $\Lambda$ in the above mentioned range (see the Appendix \ref{ap}).

(v) The theory, together with the form factor, is finite. For each value of
$m$ (and hence of the momentum $k$) the model is mathematically well defined
(no matter how large $m$ is). In fact, the normalization of the spectral
function in Eq.~(\ref{norm}) (see below) is realized by formally
integrating up to infinity. Of course, far from the energy of interest,
the model - even if mathematically well defined - cannot represent a reliable description of reality, since other resonances are missing. We thus consider
our approach valid up to $m\lesssim4$ GeV. The important point, however, is
that we do not require that $k$ should be smaller than $\Lambda$: simply, when
$k$ is larger than $\Lambda,$ that decay mode is naturally suppressed.
Similarly, in the already mentioned $^{3}P_{0}$ model, it is common to have
values of $k$ larger than $\Lambda$ \cite{amslerclose}.

 Since the theory is finite, there is no need of any
subtraction constant. Even the bare quarkonium mass $m_{0}$ should be
regarded as the mass of the charmonium state in the large-$N_{c}$
limit, which is then shifted by (finite) loop corrections. A different
approach than the one followed in this work would be to consider subtractions.
In the present case, a three-time subtracted dispersion relation would be
needed to guarantee convergence. For a detailed treatment of such an approach
as well as the subtleties of a three-time subtraction, see the analogous case
described in\ Ref. \cite{3770} (Appendix D therein).\ The results were not
satisfactory. Moreover, in Ref. \cite{3770} it is also clarified that the
model followed by us and the one with subtractions are different and that the
introduction of a low-energy scale is preferable for the type of models
considered in this work and similar ones, e.g. Refs.
\cite{nl,nl2,a0,soltysiak,boglione1,boglione2,tornqvist,milenapeter} %
.

(vi) The mixing of the bare charmonium state with meson-meson states is also part of other approaches based on quantum mechanics in which quarks d.o.f. explicitely enter, e.g. \cite{coito3872,santopinto,xiao1,xiao2}. Typically, an effective Hamiltonian describing the transtion from charmonia to $DD^*$ mesons is used, as the Friedrichs model of \cite{xiao1,xiao2}. In our approach, this mixing is described at the quantum field theoretical level (the vertices are fixed by invariances under Lorentz, parity, and charge conjugation transofrmations). Moreover, in our case no quarks d.o.f. are explicitely present, since they are hidden into the bare field $\chi_{c1}$.

\bigskip

\textit{The propagator and the spectral function}: The scalar part of the
propagator of the field $\chi_{c1}$, as function of the variable $s=m^{2}$, reads:%

\begin{equation}
\Delta(s)=\frac{1}{s-m_{0}^{2}+\Pi(s)}\text{ ,} \label{prop}%
\end{equation}
where $m_{0}\approx3.95$ GeV is the bare quark-antiquark mass of $\chi
_{c1}$ predicted by quark models \cite{isgur}. The
quantity
\begin{equation}
\Pi(s)=\Pi_{D^{\ast0}\bar{D}^{0}+h.c.}(s)+\Pi_{D^{\ast+}D^{-}+h.c.}(s)=g_{\chi
_{c1}DD^{\ast}}^{2}\left[  \Sigma_{D^{\ast0}\bar{D}^{0}+h.c.}(s)+\Sigma
_{D^{\ast+}D^{-}+h.c.}(s)\right]\   \label{sigmas}%
\end{equation}
is the self-energy contribution, which is the sum of the $D^{\ast0}\bar{D}%
^{0}$ and $D^{\ast+}D^{-}$ loops. We note that, at the one-loop level, the quantities
$\Sigma_{D^{\ast0}\bar{D}^{0}+h.c.}$ and $\Sigma_{D^{\ast+}D^{-}+h.c.}$, defined
in\ Eq.~(\ref{sigmas}), do not depend on the coupling constant $g_{\chi
_{c1}DD^{\ast}}^{2}.$ As shown in Ref.~\cite{schneitzer}, the one-loop level is
a very good approximation for hadronic phenomenology (no need to consider
diagrams in which the unstable state is exchanged by the decay products).

At one-loop, the imaginary part reads%
\begin{equation}
\operatorname{Im}\Pi(s)=\sqrt{s}\left[  \Gamma_{\chi_{c1}(2P)\rightarrow
D^{\ast0}\bar{D}^{0}+h.c.}(\sqrt{s})+\Gamma_{\chi_{c1}(2P)\rightarrow D^{\ast
+}D^{-}+h.c.}(\sqrt{s})\right]  \text{ }.
\end{equation}
The real part is obtained by the following dispersion relation (valid for
$\sqrt{s}$ real and larger than $m_{D^{\ast+}}+m_{D^{-}}$) :
\begin{align}
\operatorname{Re}\Pi(s)  &  =\frac{PP}{\pi}\int_{(m_{D^{\ast0}}+m_{D^{0}}%
)^{2}}^{\infty}\sqrt{s^{\prime}}\frac{\Gamma_{\chi_{c1}(2P)\rightarrow
D^{\ast0}\bar{D}^{0}+h.c.}(\sqrt{s^{\prime}})}{s^{\prime}-s}\ \mathrm{d}%
s^{\prime}\nonumber\\
&  +\frac{PP}{\pi}\int_{(m_{D^{*+}}+m_{D^{+}})^{2}}^{\infty}\sqrt{s^{\prime}}%
\frac{\Gamma_{\chi_{c1}(2P)\rightarrow D^{\ast+}D^{-}+h.c.}(\sqrt{s^{\prime}}%
)}{s^{\prime}-s}\ \mathrm{d}s^{\prime}.
\end{align}
For $s<\left(m_{D^{\ast0}}+m_{D^{0}}\right)
^{2}$ or having a nonzero imaginary part, $PP$ is omitted and $\Pi(s=z^{2})$, in the I Riemann sheet (RS), is:

\begin{equation}
\Pi(s=z^{2})=\frac{1}{\pi}\int_{\left(  m_{D^{\ast0}}+m_{D^{0}}\right)  ^{2}%
}^{\infty}\sqrt{s^{\prime}}\frac{\Gamma_{\chi_{c1}\rightarrow D^{\ast0}\bar
{D}^{0}+h.c.}(\sqrt{s^{\prime}})}{s^{\prime}-z^{2}}\ \mathrm{d}s^{\prime}%
+\frac{1}{\pi}\int_{\left(  m_{D^{*+}}+m_{D^{+}}\right)  ^{2}}^{\infty}%
\sqrt{s^{\prime}}\frac{\Gamma_{\chi_{c1}\rightarrow D^{\ast+}D^{-}+h.c.}%
(\sqrt{s^{\prime}})}{s^{\prime}-z^{2}}\ \mathrm{d}s^{\prime}\text{ }.
\label{selfenergycomplex}%
\end{equation}
It should be stressed that $\Pi(z^{2}\rightarrow\infty)\rightarrow0$ in all
directions of the complex plane. In order to avoid misunderstanding, we recall
that, in the I RS, the function $\Pi(z^{2}\rightarrow\infty)$ is an
utterly different complex function than $e^{-z^{2}/\Lambda^{2}}$, see Ref.~\cite{3770} for a detailed discussion of this point. In fact, in the I RS, besides the cut along the real axis, $\Pi(z^{2})$ is regular and well defined
everywhere and does not contain any singular point, contrary to $e^{-z^{2}%
/\Lambda^{2}},$ which contains an essential singularity at $\infty.$ In other
Riemann sheets, the properties of $\Pi(z^{2})$ are different and singular points
are present.

The spectral function (or the mass distribution) is defined as
\begin{equation}
d_{\chi_{c1}(2P)}(m)=-\frac{2m}{\pi}\operatorname{Im}[\Delta(s=m^{2})]\text{
.} \label{dchi}%
\end{equation}
The quantity $\mathrm{dm}d_{\chi_{c1}(2P)}(m)$ represents the probability that
the unstable state has a mass between $m$ and $m+\mathrm{dm}$
\cite{salam,lupo,lupoprd,duecan}. It fulfills the important normalization
condition
\begin{equation}
\int_{m_{D^{\ast0}}+m_{D^{0}}}^{\infty}\mathrm{dm}\ d_{\chi_{c1}(2P)}(m)=1\text{
.} \label{norm}%
\end{equation}
This is a consequence of the K\"{a}ll\'{e}n--Lehmann representation and of
unitarity. For a rigorous proof, using the vertex function regularization (and
for the link to other regularization schemes), see Ref.~\cite{lupoprd}. In our
approach, the normalization follows automatically from the formalism. The numerical verification of Eq.~(\ref{norm}) represents an important check of the correctness of the numerically performed calculations.

\bigskip

\textit{The definition(s) of the mass(es): }An unstable state is described by
its mass distribution and, strictly speaking, it does not have a definite mass.
Nevertheless, one can define it in various ways. A typical one is the so-called Breit-Wigner (BW) mass, given by
$\operatorname{Re}[\Delta^{-1}(s=m_{BW}^{2})]=0$. This equation is however
meaningful when a unique and symmetric peak of the spectral function is
present. As we shall see, this is not true in our case. Alternatively, it is
common to search for the position(s) of the pole(s) in the complex plane
$\Delta^{-1}(s=z_{pole}^{2})=0$ and then identify the mass as its real part,
$m_{pole}=\operatorname{Re}[z_{pole}]$ (in the proper Riemann sheet).
Here, this second approach is more useful for both poles, the standard
$\bar{c}c$ one, in the III RS, and the virtual one, linked to $X(3872)$, on the II RS.

\bigskip

\textit{Radiative decays}: Besides the dominant decay into $DD^{\ast}$, the
terms that describe the radiative decays read:%

\begin{equation}
\mathcal{L}_{\chi_{c1}\text{-rad}}=g_{\chi_{_{c1}}\psi(1S)\gamma}\ %
\chi_{c1,\mu}\psi(1S)_{\nu}\tilde{F}^{\mu\nu}+g_{\chi_{_{c1}}\psi
(2S)\gamma}\ \chi_{c1,\mu}\psi(2S)_{\nu}\tilde{F}^{\mu\nu}+...\ \text{,}
\label{radlag}%
\end{equation}
where the coupling constants $g_{\chi_{_{c1}}\psi(1S)\gamma}$ and
$g_{\chi_{_{c1}}\psi(2S)\gamma}$ can be determined by the quark model
\cite{barnesgodfrey} (through the overlap of wave functions). Later on, we shall use the
couplings determined in\ Ref.~\cite{barnesgodfrey}.

\section{\label{s3}Results}

In this Section, we present the results by starting from the spectral
functions, poles, and strong decays (Sec.~\ref{s3a}). Later on, we focus on
radiative decays, prompt production, and isospin breaking decays (Sec.~\ref{s3b}).

\subsection{\label{s3a}Spectral function, poles, and strong decays}

\textit{Case I (}$m_{0}=3.95$ GeV)\textit{:} As a first case, let us set
the bare mass to $m_{0}=3.95$ GeV (close to the value predicted by the quark
model in Ref.~\cite{isgur}). For the parameter $\Lambda$ we shall use $0.5$ GeV (this
is a typical value for $\Lambda;$ later, we will check the dependence on
$\Lambda$). The masses of the pseudoscalar states read $m_{D^{0}}=1.86483$ GeV
and $m_{D^{+}}=1.86959$ GeV, and the masses of the vector states are
$m_{D^{\ast0}}=2.00685$ GeV and $m_{D^{\ast+}}=2.01026$ GeV. Hence, the
relevant thresholds are: $m_{D^{0}}+m_{D^{\ast0}%
}=3.87168$ GeV and $m_{D^{+}}+m_{D^{\ast+}}=3.87985$ GeV.

We determine the coupling constant $g_{\chi_{c1}DD^{\ast}}=9.732$ GeV by
requiring that:
\begin{equation}
\operatorname{Re}[\Delta^{-1}(s=m_{\ast}^{2})]=0, \text{ for }m_{\ast
}=3.874\text{ GeV .}\label{mstar}%
\end{equation}
The chosen value of $m_{\ast}$ is slightly \textit{above} the $D^{\ast0}%
D^{0}$ threshold, but below the $D^{\ast+}D^{-}$ one. Again, the precise
value of $m_{\ast}$ is not so important, as long as it is between the two
thresholds (see below).

The spectral function $d_{\chi_{c1}(2P)}(m)$ is plotted in Fig.~\ref{f1}. It has a
very peculiar form: a extremely narrow and high peak \textit{very} close to the
$D^{\ast0}D^{0}$ threshold is realized. We identify this peak with the resonance $X(3872).$ A second and broad
peak at $3.986$ GeV is also visible, and it corresponds to a roughly $80$ MeV
broad state, in agreement with old and recent quark model predictions. It should be however stressed that
the whole spectral function of Fig.~\ref{f1}, even if it contains two peaks,
originates from one single seed state, and it is correctly normalized to 1, as we
verify numerically upon integrating to $10$ GeV (much larger than the energy
scale involved, hence \textit{de facto} infinity). This in turn shows that,
as anticipated, the model is mathematically consistent up to large values.

Between the thresholds one has:
\begin{equation}
\int_{m_{D^{\ast0}}+m_{D^{0}}}^{m_{D^{+}}+m_{D^{\ast+}}}\mathrm{dm}\ %
d_{\chi_{c1}(2P)}(m)=0.160\text{ ,}%
\end{equation}
thus $16\%$ of the whole spectral function is contained in this energy interval. Yet, since the $X(3872)$ is definitely narrower than $8$ MeV, and the
experimental uncertainty of its width is roughly $1$ MeV, we also consider the integral%
\begin{equation}
\int_{m_{D^{\ast0}}+m_{D^{0}}}^{m_{D^{\ast0}}+m_{D^{0}}+1\text{ MeV}%
}\mathrm{dm}\ d_{\chi_{c1}(2P)}(m)=0.049\text{ ,} \label{integralthr}%
\end{equation}
that we interpret as it follows: the $X(3872)$ corresponds roughly to $4.9\%$ of
the whole object described by $d_{\chi_{c1}(2P)}(m).$ In other parts of the work, we shall repeat the $`1$
MeV' estimate for the extension of the peak associated to the $X(3872)$.

In Fig.~\ref{f2}, we plot the function $\operatorname{Re}[\Delta^{-1}(s=m^{2})]$: by
construction, it has a zero at $m_{\ast}=3.874$ GeV, which is responsible for the high
peak associated with the $X(3872)$, at the $D_{0}D_{0}^{\ast}$ threshold. 
There is another zero at $m=3.973$ GeV, corresponding to the broad peak on the
right. A
third zero, at $3.891$ GeV, does not correspond to any peak, since the
derivative of the function is negative (see the discussion in Ref.~\cite{boglione1}).

As expected, there is a pole on the III RS, which comes from the $\bar{c}c$ seed state, that relates to the broad peak:
\begin{equation}
3.995-i0.036 \text{ GeV} \text{ .} \label{seedpole1}%
\end{equation}
Thus, a pole width of $72$ MeV follows. In addition, a virtual state is
obtained: there is a pole on the II RS on the real axis, just below
the $D^{\ast0}D^{0}$ threshold, for:%
\begin{equation}
3.87164-i \varepsilon \text{ GeV.} \label{virtualpole}%
\end{equation}
This is the pole associated to $X(3872),$ appearing as a narrow peak above
the threshold in Fig.~\ref{f1}.

Summarizing:
\begin{equation}
X(3872)\Leftrightarrow\left\{
\begin{array}
[c]{c}%
\text{zero of }\operatorname{Re}[\Delta^{-1}(s=m^{2})]\text{ for }m_{\ast
}=3.874\ \mathrm{GeV}\\
\text{virtual pole on the II RS for }3.87164-i\varepsilon\text{ GeV}%
\end{array}
\right.  \text{ .}%
\end{equation}
Note, the virtual pole is just {$0.04$ MeV below the $D^{0}D^{\ast 0}$
threshold. Of course, the precise value of $m_{\ast}$ and the virtual pole
vary when changing the parameters, but the overall picture is quite stable.%

\begin{figure}[h!]
\begin{center}
\includegraphics[width=0.8 \textwidth]{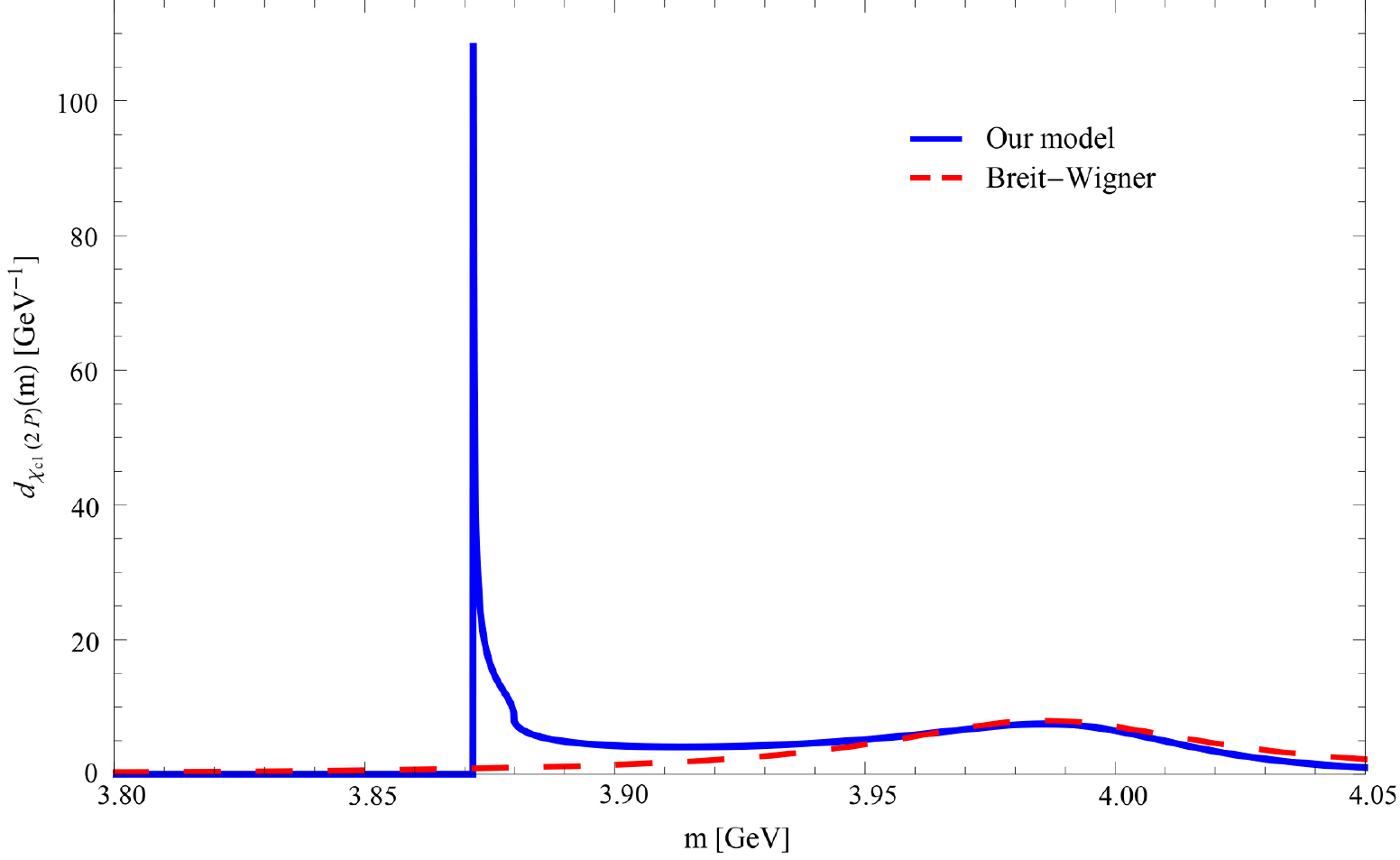}
\caption{\label{f1}Solid line: spectral function $d_{\chi_{c1}(2P)}(m)$ in
Eq.~(\ref{dchi}), for the case I (see text). Two peaks are present, linked to two distinct poles. A broad peak, originated by the seed $\bar{c}c$ state, at about $3.99$ GeV is present on the right,
while a narrow and high peak is located just on the right of the lowest threshold: it results from $DD^{\ast}$ loops dressing the $\bar{c}c$, and corresponds to the well
known state $X(3872).$ The dashed line corresponds to a Breit-Wigner
approximation for the seed state, with parameters $m_{BW}=3.986$ GeV and
$\Gamma_{BW}=79.7$ MeV (cf.~Eqs.~(\ref{wch1}) and (\ref{wch2})).}
\end{center}
\end{figure}

\bigskip%

\begin{figure}[h!]
\begin{center}
\includegraphics[width=0.8 \textwidth]{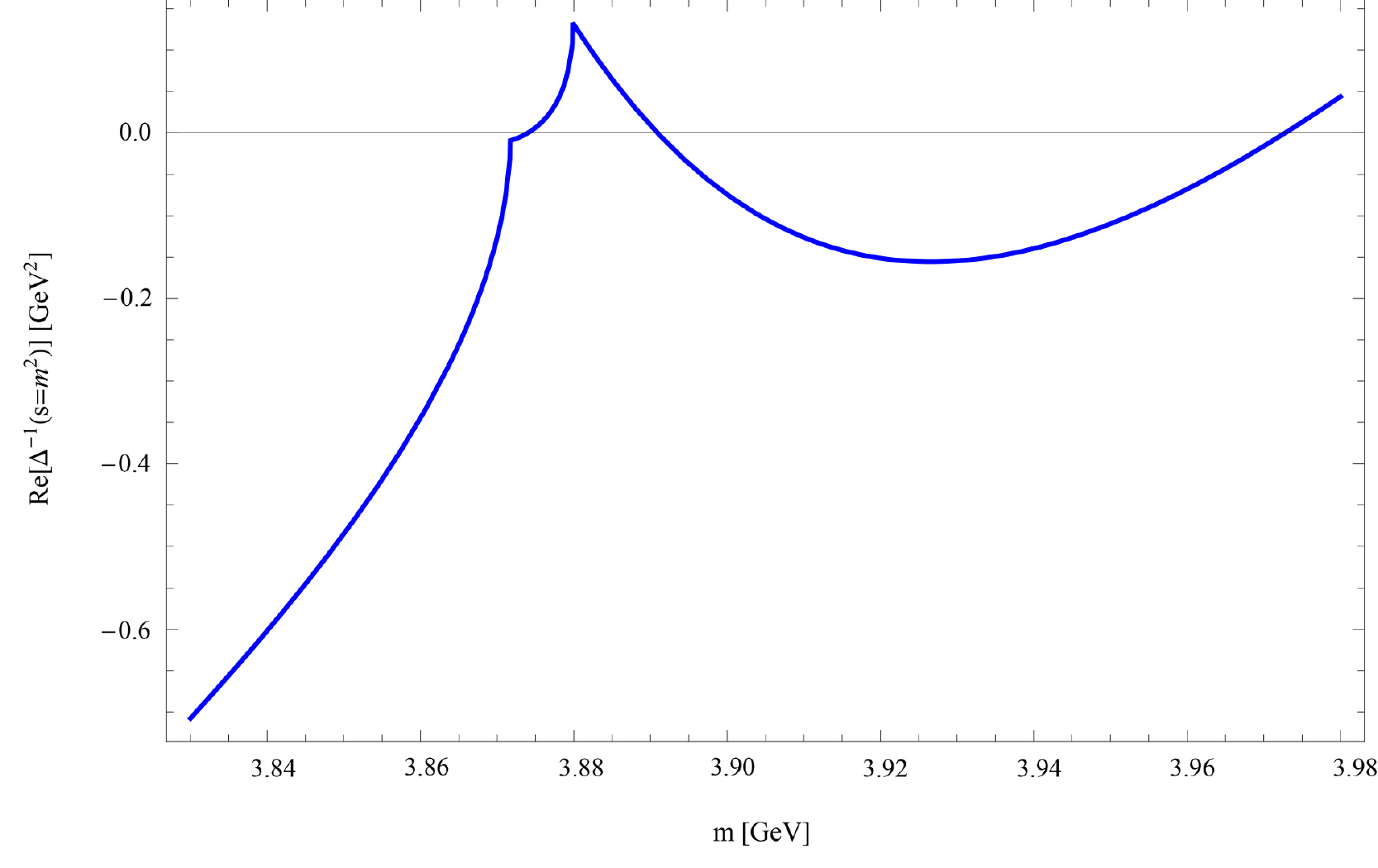}
\caption{\label{f2}Plot of the function $\operatorname{Re}[\Delta^{-1}(m^{2})]$ (see Eq.~(\ref{prop})) for case I (see text). Three zeros are present: on the right for
$m=3.973$ GeV, which corresponds roughly to the broad peak of the spectral function,
and to the pole in the III RS of the `seed' charmonium $2$ $^{3}P_{1}$ state;
on the left for $m_{\ast}=3.874$ GeV (in between the thresholds), which
corresponds to the narrow peak at the $D^{0}D^{\ast 0}$ threshold and to a
virtual pole on the II RS; the zero in the middle does not lead to peaks or
poles.}
\end{center}
\end{figure}

Next, we turn to decay widths. From Fig.~\ref{f1}, it is clear that the spectral function has not a Breit-Wigner form. Nevertheless, for the maximal height of the broad peak, which is located at $3.986$ GeV, reasonable estimates are found:
\begin{align}
\Gamma_{\chi_{c1}(2P)\rightarrow D^{\ast0}\bar{D}^{0}+h.c.}(3.986\text{ GeV}) &
=38.1\text{ MeV ,}\label{wch1}\\
\Gamma_{\chi_{c1}(2P)\rightarrow D^{\ast+}D^{-}+h.c.}(3.986\text{ GeV}) &
=41.6\text{ MeV ,}\label{wch2}%
\end{align}
for a total width of $79.7$ MeV: this is the width used for the Breit-Wigner function in Fig.~\ref{f1} (note, using the pole mass would generate similar results). 

For the peak close to threshold, identified with $X(3872)$, a direct evaluation of
the width at the peak is not really useful. The peak itself has a very small
width at half height ($\sim0.1$ MeV). Even when changing the parameters, it is
always smaller than $0.5$ MeV. As a good estimate of the dominant decay of the
$X(3872)$, we consider the following average value, which extends from the 
threshold to the left-threshold plus $1$ MeV, that includes the peak:
\begin{equation}
\Gamma_{X(3872)\rightarrow D^{\ast0}\bar{D}^{0}+h.c.}^{\text{average}}%
=\int_{m_{D^{\ast0}}+m_{D^{0}}}^{m_{D^{\ast0}}+m_{D^{0}}+1\text{ MeV}%
}\mathrm{dm}\ \Gamma_{\chi_{c1}(2P)\rightarrow D^{\ast0}\bar{D}^{0}%
+h.c.}(m)d_{\chi_{c1}(2P)}(m)=0.61\text{ MeV. } \label{intwidth}%
\end{equation}
Thus, the model shows that the decay of $X(3872)$ into $D^{\ast0}\bar{D}%
^{0}+h.c.$ is sizable: in fact, $0.61$ MeV is comparable to the maximal value
of about $1$ MeV, estimated for the $X(3872).$ On the other hand, $\Gamma
_{X(3872)\rightarrow D^{\ast+}D^{-}+h.c.}$ vanishes because $X(3872)$ is always
subthreshold for this decay channel.

Note, in this work we use a relativistic treatment, which naturally appears
in our QFT framework previously developed for low-energy mesons. Actually,
relativistic effects are not expected to affect much the left part of the
line-shape of Fig. 1 (and later on of Fig. 3) since it corresponds to a pole
which is very close to the left threshold, but could be potentially relevant
for the seed state, which is $80$ MeV broad and quite far from it. In this
respect, our relativistic treatment may be useful for a detailed understanding
of the system. 

In the Appendix \ref{ap}, we discuss the results for other parameter choices, for which there are no big qualitative changes: the overall picture is
quite stable. The variation of the results also represents an estimate of the
uncertainties of our analysis. In \ref{ap2}, we repeat
the study for different values of $\Lambda$, and keep $m_{\ast}=3.874$ GeV fixed by adjusting the coupling constant. On the other hand, in \ref{ap1} we change the coupling
constant $g_{\chi_{c1}DD^{\ast}}$, and keep $\Lambda$ fixed by varying $m_*$. For smaller couplings, $m_{\ast}$ moves to the right threshold $m_{D^{+}}+m_{D^{\ast+}}$, and the height of the peak $X(3872)$ decreases and gradually fades away. When $g_{\chi_{c1}DD^{\ast}}$ becomes too small, $m_{\ast}$ exceeds $m_{D^{+}}+m_{D^{\ast+}}$ and the peak associated to the $X(3872)$ disappears. When $g_{\chi_{c1}DD^{\ast}}$ increases, $m_{\ast}$ moves to the left towards the $D^{\ast0}D^{0}$ threshold, and the height of the 
$X(3872)$ increases. For coupling constants exceeding the critical value
$g_{\chi_{c1}DD^{\ast}}^{\text{critical}}=9.808$ GeV, for which $m_{\ast
}$ lies just at the $m_{D^{\ast0}}+m_{D^{0}}$ threshold, there is a pole in the
first Riemann sheet: an additional (quasi-)stable bound state emerges.\ In such case, the spectral function takes the form \cite{lupo}:%
\begin{equation}
d_{\chi_{c1}(2P)}(m)=Z\delta(m-m_{BS})+d_{\chi_{c1}(2P)}^{\text{above
threshold}}(m)\label{realpoleds}%
\end{equation}
where $m_{BS}$ is the mass of the bound-state, to be interpreted as a
dynamically generated molecular-like state (still, this bound state is deeply
connected to the seed state and cannot exist without it). The normalization
(\ref{norm}) is still formally valid, leading to:
\begin{equation}
Z+\int_{m_{D^{\ast0}}+m_{D^{0}}}^{\infty}\mathrm{dm}\ d_{\chi_{c1}%
(2P)}^{\text{above threshold}}(m)=1.
\end{equation}
For instance, for $g_{\chi_{c1}DD^{\ast}}=10$ GeV (just above the critical
value), one has $Z=0.0465$ and $m_{BS}=3.87164$ GeV (hence, this is a pole on
the I RS). The shape of $d_{\chi_{c1}(2P)}^{\text{above
threshold}}$ is very similar to Fig.~\ref{f1}, with the important difference that the
area does not sum up to unity. It is then very difficult to distinguish the case in Fig.~\ref{f1} from  this latter one, even if there is an important difference: {\it virtual} versus {\it real} pole. A (quasi-)bound state, below threshold, neither decays into $D^{\ast0}D^{0}$ nor into $D^{\ast+}D^{-}$, but only into suppressed radiative and light hadron decays, thus the width associated to this
pole is very small, but the peak just above the $D^{0}D^{\ast0}$ threshold is
still present. Note, the integral of the function $d_{\chi_{c1}(2P)}^{\text{above
threshold}}(m)$, between the two thresholds, amounts to $0.133$, while between
$m_{D^{\ast0}}+m_{D^{0}}$ and $m_{D^{\ast0}}+m_{D^{0}}+1$ MeV to $0.035$, thus slightly
smaller then the previous case reported in Eq.~\eqref{integralthr}.

\bigskip

\textit{Case II (}$m_{0}=3.92$ GeV)\textit{:} We repeat the study for a different value of the bare mass $m_{0}$. We use $m_{0}=3.92$ GeV, slightly
smaller than the value in Ref.~\cite{isgur} and close to the value in Ref.~\cite{ebert}. We determine the coupling constant
$g_{\chi_{c1}DD^{\ast}}=7.557$ GeV by requiring, as before, that $m_{\ast
}=3.874$ GeV. The spectral function is depicted in Fig.~\ref{f3}. There is still a
very pronounced peak close to the $D^{0}D^{\ast0}$ threshold, but there is no
peak at higher values: the broad peak corresponding to the seed melts with the whole structure.

The amount of spectral function between the thresholds is:%

\begin{equation}
\int_{m_{D^{\ast0}}+m_{D^{0}}}^{m_{D^{+}}+m_{D^{\ast+}}}\mathrm{dm}\ d_{\chi_{c1}%
(2P)}(m)=0.250\text{ ,}%
\end{equation}
while the one associated to the $X(3872)$ is
\begin{equation}
\int_{m_{D^{\ast0}}+m_{D^{0}}}^{m_{D^{\ast0}}+m_{D^{0}}+1\text{ MeV}}%
\mathrm{dm}\ d_{\chi_{c1}(2P)}(m)=0.067\text{ .}%
\end{equation}
In Fig.~\ref{f4} we also plot $\operatorname{Re}[\Delta^{-1}(s=m^{2})].$ As
required, there is one zero at $3.874$ GeV. However, no other intersection is
present: this is in agreement with the absence of the broad peak in Fig.~\ref{f3}. Even
if there is no peak, there is indeed a pole on the III Riemann sheet (seed
state):
\begin{equation}
3.953-i0.044\text{ GeV ,}%
\end{equation}
pretty similar to the case I presented in Eq. (\ref{seedpole1}). Thus, a pole
width of $88$ MeV follows. This example shows how important it is to look for
poles: a state may exist even when there is no bump. In
addition, a virtual state on the II RS\ is obtained for:%
\begin{equation}
3.87160-i\varepsilon\text{ GeV,}%
\end{equation}
just $0.08$ MeV below the $\bar{D}^{0}D^{\ast 0}$ threshold.

Let us turn to the partial widths. Since there is no peak, we use the pole
value of $3.953$ GeV for the on-shell mass:
\begin{align}
\Gamma_{\chi_{c1}(2P)\rightarrow D^{\ast0}\bar{D}^{0}+h.c.}(3.953)  &
=32.9\text{ MeV}\text{ },\label{sd1}\\
\Gamma_{\chi_{c1}(2P)\rightarrow D^{\ast+}D^{-}+h.c.}(3.953)  &  =35.4\text{
MeV}\text{ }. \label{sd2}%
\end{align}
The peak close to threshold, denoted as $X(3872)$, has a
width at half height of $\sim 0.91$ MeV. The integrated signal
\begin{equation}
\Gamma_{X(3872)\rightarrow D^{\ast0}\bar{D}^{0}+h.c.}^{\text{average}}%
=\int_{m_{D^{\ast0}}+m_{D^{0}}}^{m_{D^{\ast0}}+m_{D^{0}}+1\text{ MeV}%
}\mathrm{dm}\ \Gamma_{\chi_{c1}(2P)\rightarrow D^{\ast0}\bar{D}^{0}%
+h.c.}(m)d_{\chi_{c1}(2P)}(m)=0.54\text{ MeV }%
\end{equation}
is very similar to case I. This is in general a quite stable
outcome of our study.

Increasing and decreasing the cutoff $\Lambda$ and the coupling constant
$g_{\chi_{c1}DD^{\ast}}$ generates the same type of changes mentioned above,
see the Appendix \ref{ap} for further details.

Finally, for related studies in which the spectral function plays a major
role, we refer to e.g. Refs. \cite{takeuchi,kalashnikova2,braatenlineshape},
where in a non-relativistic context a shape of the spectral function
qualitative similar to our Fig. 1 was obtained. Yet, the outcome of our Fig. 3
(two poles but only one peak) seems to be a peculiar outcome of our work.'

\begin{figure}[h!]
\begin{center}
\includegraphics[width=0.8 \textwidth]{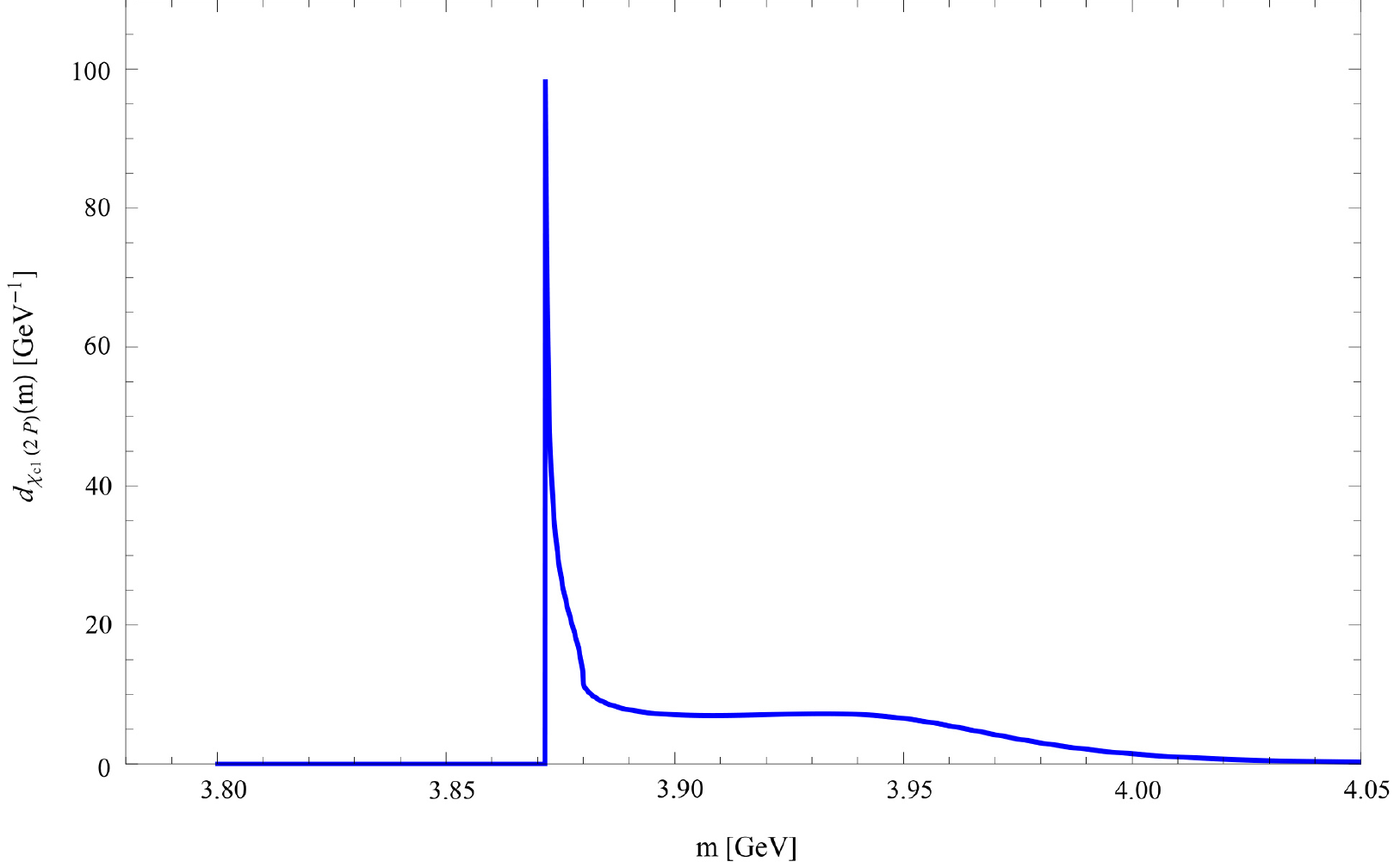}
\caption{\label{f3}Solid line: spectral function for case II (see text). As it is
visible, there is no peak for the seed state, but the pronounced
peak corresponding to $X(3872)$ still exists.}
\end{center}
\end{figure}


\begin{figure}[h!]
\begin{center}
\includegraphics[width=0.8 \textwidth]{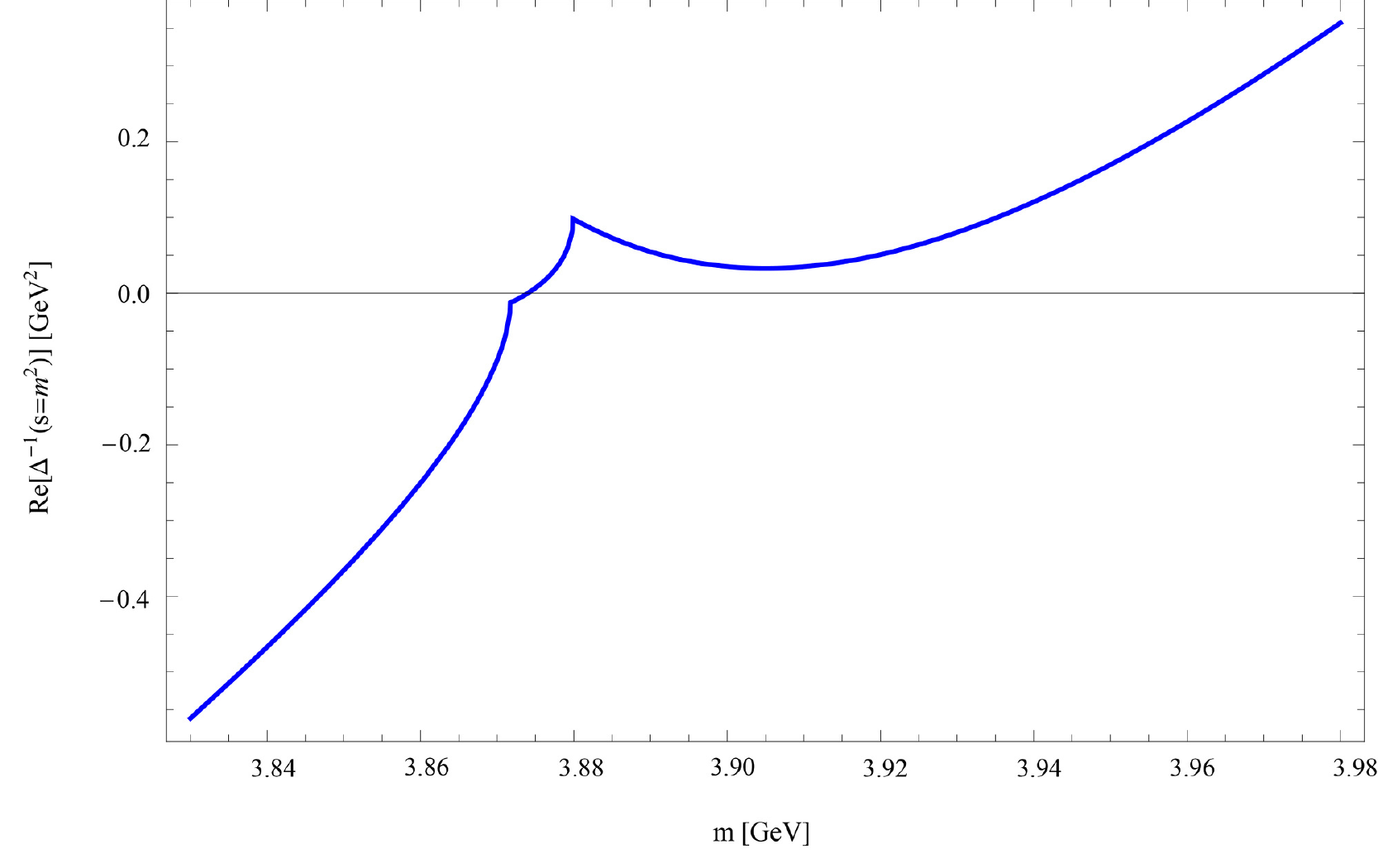}
\caption{\label{f4}Plot of the function $\operatorname{Re}[\Delta^{-1}(m^{2})]$ (see Eq.~(\ref{prop})) for case II (see text). Only one zero at $m_{\ast}=3.874$ GeV
(in between the thresholds) is present, which corresponds to the narrow peak
at the $D^{0}D^{\ast 0}$ threshold and to a virtual pole on the II RS.}
\end{center}
\end{figure}

\subsection{\label{s3b}Consequences of the approach}

Next, we study various consequences of the approach: radiative decays, prompt
production, and, most notably, the isospin-breaking strong decay. We use the
numerical values of case I in Sec.~\ref{s3a}. For other parameter choices, see
the Appendix \ref{ap}.

\bigskip

\textit{Radiative decays}: The coupling constants 
$g_{\chi_{_{c1}}%
(2P)\psi(1S)\gamma}$ and $g_{\chi_{_{c1}}(2P)\psi(2S)\gamma},$ entering Eq.~(\ref{radlag}) and describing the
radiative transitions $\chi_{_{c1}}(2P)\rightarrow\psi(1S)\gamma$ and
$\chi_{_{c1}}(2P)\rightarrow\psi(2S)\gamma$, can be calculated within the
quark model. They are proportional to the overlap of the spatial wave
functions of $\chi_{_{c1}}(2P)$ with, respectively, $\psi(1S)$ and $\psi(2S).$
It follows that $g_{\chi_{_{c1}}(2P)\psi(2S)\gamma}$ is larger than
$g_{\chi_{_{c1}}(2P)\psi(1S)\gamma}$, since the $2P\rightarrow2S$ overlap is
larger than $2P\rightarrow1S$ (this is due to the fact that in the latter a
cancellation due to the node, present only in $2P$, occurs).

The decay widths of the processes $\chi_{_{c1}}(2P)\rightarrow\psi(1S)\gamma$
and $\chi_{_{c1}}(2P)\rightarrow\psi(2S)\gamma$, as a function of the running
mass $m$ of $\chi_{_{c1}}(2P)$, read:%
\begin{align}
\Gamma_{\chi_{_{c1}}(2P)\rightarrow\psi(1S)\gamma}(m)  &  =g_{\chi_{_{c1}%
}(2P)\psi(1S)\gamma}^{2}\frac{k^{3}(m,m_{\psi(1S)},0)}{8\pi m^{2}}\frac{4}%
{3}\left(  1+\frac{k^{2}(m,m_{\psi(1S)},0)}{m_{\psi(1S)}^{2}}\right)  \text{
,}\\
\Gamma_{\chi_{_{c1}}(2P)\rightarrow\psi(2S)\gamma}(m)  &  =g_{\chi_{_{c1}%
}(2P)\psi(2S)\gamma}^{2}\frac{k^{3}(m,m_{\psi(2S)},0)}{8\pi m^{2}}\frac{4}%
{3}\left(  1+\frac{k^{2}(m,m_{\psi(2S)},0)}{m_{\psi(2S)}^{2}}\right)  \text{
.}%
\end{align}
Hence, their ratio is%

\begin{equation}
\frac{\Gamma_{\chi_{_{c1}}(2P)\rightarrow\psi(2S)\gamma}(m)}{\Gamma
_{\chi_{_{c1}}(2P)\rightarrow\psi(1S)\gamma}(m)}=\left(  \frac{g_{\chi_{_{c1}%
}(2P)\psi(2S)\gamma}}{g_{\chi_{_{c1}}(2P)\psi(1S)\gamma}}\right)  ^{2}\left(
\frac{k(m,m_{\psi(2S)},0)}{k(m,m_{\psi(1S)},0)}\right)  ^{3}\left(
\frac{1+\frac{k^{2}(m,m_{\psi(2S)},0)}{m_{\psi(2S)}^{2}}}{1+\frac
{k^{2}(m,m_{\psi(1S)},0)}{m_{\psi(1S)}^{2}}}\right)  \text{ }.\label{radratio}%
\end{equation}

Following Ref.~\cite{barnesgodfrey}, $g_{\chi_{_{c1}}(2P)\psi(2S)\gamma}\propto\sqrt
{\alpha_{QED}}\left\langle 2S\left\vert r\right\vert 2P\right\rangle $ and
$g_{\chi_{_{c1}}(2P)\psi(1S)\gamma}\propto\sqrt{\alpha_{QED}}\left\langle
1S\left\vert r\right\vert 2P\right\rangle $, where the numerical values are
$\left\langle 2S\left\vert r\right\vert 2P\right\rangle \simeq2.72$ GeV$^{-1}$
and $\left\langle 1S\left\vert r\right\vert 2P\right\rangle \simeq0.15$
GeV$^{-1}$ \cite{barnesgodfrey}. We recall that $\chi_{_{c1}}(2P)$ in Eq.~(\ref{radratio}) refers to the whole energy domain of the state. For the decays of $X(3872)$, we set $m\simeq m_{X(3872)}%
\simeq 3.872$ GeV: we obtain $\frac{\Gamma_{X(3872)\rightarrow\psi(2S)\gamma}%
}{\Gamma_{X(3872)\rightarrow\psi(1S)\gamma}}\simeq 5.4$. This is
comparable to (but somewhat larger than) the experimental value $2.6\pm0.4$
reported by the PDG \cite{pdg}.\ We also recall the values $2.38\pm0.64\pm0.29$,
determined by the LHCb collaboration \cite{lhcbgamma} and $3.4\pm1.4$,
determined by the BABAR collaboration \cite{babargamma} (see also the
theoretical discussion in Ref.~\cite{simonov}).

The main point concerning the radiative decays is that our approach naturally
explains the large $\psi(2S)\gamma$ to $\psi(1S)\gamma$ ratio, since the
$\bar{c}c$ component provides the dominant contribution to these decays, see Eq.~(\ref{radlag}). Thus, this feature applies for both
the seed state and for $X(3872).$ On the contrary, a purely molecular state would
deliver the opposite result: $\left\langle 2S\left\vert r\right\vert
2P\right\rangle $ much smaller than $\left\langle 1S\left\vert r\right\vert
2P\right\rangle .$ Yet, as discussed in Ref.~\cite{guo}, a $DD^{\ast}$
component is not excluded. (Indeed, our sligth overestimation of the ratio can be caused by not considering  $DD^{\ast}$ loop processes in a way similar to the isospin breaking decays described below). 

Finally, we use our approach to estimate the decay widths for the two channels,
upon integrating over the spectral function:
\begin{align}
\Gamma_{X(3872)\rightarrow\psi(1S)\gamma}  &  =\int_{m_{D^{\ast0}}+m_{D^{0}}%
}^{m_{D^{\ast0}}+m_{D^{0}}+1\text{ MeV}}\mathrm{dm}\ \Gamma_{\chi_{_{c1}%
}(2P)\rightarrow\psi(1S)\gamma}(m)d_{\chi_{c1}(2P)}(m)\simeq0.54\text{
keV\ ,}\label{raddec1}\\
\Gamma_{X(3872)\rightarrow\psi(2S)\gamma}  &  =\int_{m_{D^{\ast0}}+m_{D^{0}}%
}^{m_{D^{\ast0}}+m_{D^{0}}+1\text{ MeV}}\mathrm{dm}\ \Gamma_{\chi_{_{c1}%
}(2P)\rightarrow\psi(2S)\gamma}(m)d_{\chi_{c1}(2P)}(m)\simeq3.13\text{ keV\ ,}
\label{raddec2}%
\end{align}
where we have used the coupling constants $g_{\chi_{_{c1}}(2P)\psi(2S)\gamma
}=1.737$ and $g_{\chi_{_{c1}}(2P)\psi(1S)\gamma}=0.093$, extracted from Ref.~\cite{barnesgodfrey}. These widths are predictions of our approach for the radiative decays.

Some important comments about radiative decays are in order: in Ref. \cite{barnesgodfrey} the
decays -upon using the very same coupling constants that we employ here- are
larger and read 11 keV and 64 keV, respectively. \ Results of the same order
but even somewhat larger were obtained in Refs. \cite{simonov, badalian2015, wang2011}, in
which the $X(3872)$ is assumed to be a quarkonium state. It is easy to
understand why the radiative decays are smaller within our approach: since only a part
of the whole spectral function (which amounts to about $5\%$) corresponds to
the $X(3872)$, see Fig. 1, then the corresponding decays are sizably reduced.
Quite interestingly, in Ref. \cite{takeuchi2016} similar (and small) results for
the radiative decays are obtained in a hybrid model, in which the $X(3872)$ is
an admixture of molecular and charmonium components. In the purely molecular framework, the results depend on the details of the employed model and vary in a quite
wide range between 0.1-50 keV, see Refs. \cite{kalashnikova, guo} and refs. therein. Indeed,
small decay rates in the molecular approach were found in\ Ref. \cite{swanson2004}.
Presently, the experimental status for the full radiative widths is still
unclear: the combination of the upper limits $\gamma\psi(2S)/\Gamma
_{total}>0.04$ and $\gamma\psi(1S)/\Gamma_{total}>0.007$ with the lower limit
$\Gamma_{total}<1.2$ MeV \cite{pdg} does not allow for a clear statement. In the
future, new experimental results would be very useful. 

In the end, it should be stressed that our predictions take into account only
the quark-antiquark core sitting into $X(3872)$ and therefore should be
regarded as `qualitative', since other potentially relevant effects are not
yet taken into account. For instance, the decays into $\gamma\psi(1S)$ and
$\gamma\psi(2S)$ do not take place solely by the contribution of the
quark-antiquark core, but can result by the $DD^{\ast}$ loop which convert
into $J/\psi\omega$ and $J/\psi\rho$ (see later on), where the $\omega$ and
the $\rho$ further transform into $\gamma$ through the so-called vector meson
dominance, see e.g. Ref. \cite{Oconnell1997}. In this respect, the fact that our
theoretical ratio $\Gamma_{X(3872)\rightarrow\psi(2S)\gamma}/\Gamma
_{X(3872)\rightarrow\psi(1S)\gamma}$ overestimates the PDG average $2.6\pm0.6$
of a factor $2.2$ is and indication that, even if the decay into $\gamma
\psi(2S)$ is predicted to be larger than $\gamma\psi(2S)$ as it should,
improvement is needed for a more quantitative agreement in the
future.
\bigskip

\textit{Prompt production: }Within our interpretation of the $X(3872)$, its
production in heavy ion collisions can be easily explained. The reason is that 
the $\bar{c}c$ system, dressed by $DD^{\ast}$ clouds, can be regarded as a
single object described by the whole spectral function in Fig.~\ref{f1}, even if it leads to two poles, and thus two states. As seen for
radiative decays, whatever the bare coupling to the state $\chi_{1}(2P)$ is,
the very same coupling holds in general for the broad peak coming from
the seed and also for the narrow peak corresponding to the $X(3872).$

Note, a different question is the role of this resonance in thermal models. In
agreement with the case studied of Ref. \cite{begun}, the resonance $X(3872)$
is expected to be subleading \cite{arriola}. (For the appropriate theoretical framework,
see e.g.~Refs.~\cite{pok,bazak,pokfra} and refs.~therein.)

\bigskip

\textit{Isospin breaking decay}: The decay $\chi_{1}(2P)\rightarrow
\psi(1S)\omega\rightarrow\psi(1S)\pi^{+}\pi^{-}\pi^{0}$ can take place via two
different mechanisms: the first involves the emission of two gluons, which
then convert into an $\omega,$ while the second involves the
$DD^{\ast}$ loops which couple to $\psi(1S)\omega.$

On the contrary, at a first sight, the decay $\chi_{1}(2P)\rightarrow
\psi(1S)\rho^{0}\rightarrow\psi(1S)\pi^{+}\pi^{-}$ cannot occur, since
it violates isospin. The two-gluon mechanism is not possible, since two gluons
cannot convert into a $\rho$ meson. Yet, the $DD^{\ast}$ loop
generates an isospin-suppressed coupling of $\chi_{1}(2P)$ to $\psi(1S)\rho,$
as we shall discuss below.

In order to show these aspects, let us consider the following Lagrangian
coupling $DD^{\ast}$ to $\omega$ and $\rho^{0}$:
\begin{equation}
\mathcal{L}_{DD^{\ast}}=\xi_{0}D^{\ast0\mu}\bar{D}^{0}\psi(1S)^{\nu}\left[
\tilde{\omega}_{\mu\nu}+\tilde{\rho}_{\mu\nu}^{0}\right]  +\xi_{0}D^{\ast+\mu
}D^{-}\psi(1S)^{\nu}\left[  \tilde{\omega}_{\mu\nu}-\tilde{\rho}_{\mu\nu}%
^{0}\right]  +h.c.\ ,\label{DDjor}%
\end{equation}
where $\tilde{\omega}_{\mu\nu}+\tilde{\rho}_{\mu\nu}^{0}$ is proportional to
$\bar{u}u$ (and hence couples to $D^{0}\bar{D}^{0}$), while $\tilde{\omega
}_{\mu\nu}-\tilde{\rho}_{\mu\nu}^{0}$ to $\bar{d}d$ (and couples to
$D^{\ast+}D^{-}$). The constant $\xi_{0}$ is an unknown coupling constant
describing these transitions. Note, using the same $\xi_{0}$ in front of both
terms means that the Lagrangian $\mathcal{L}_{DD^{\ast}}$ fulfills isospin
symmetry. For a similar four-body interacting Lagrangian, see Ref. \cite{gamermann2}. The resulting energy dependent coupling of $\chi_{_{c1}}(2P)$ to
$\psi(1S)\rho$ is proportional to
\begin{equation}
\xi_{\chi_{_{c1}}(2P)\rightarrow\psi(1S)\rho}(m)=\xi_{0}g_{\chi_{c1}DD^{\ast}%
}\left[  \Sigma_{D^{\ast0}\bar{D}^{0}+h.c.}(s)-\Sigma_{D^{\ast+}D^{-}%
+h.c.}(s)\right]\ ,  \label{psi1rho}%
\end{equation}
while the coupling to $\psi(1S)\omega$ to%
\begin{equation}
\xi_{\chi_{_{c1}}(2P)\rightarrow\psi(1S)\omega}(m)=\xi_{0}g_{\chi_{c1}%
DD^{\ast}}\left[  \Sigma_{D^{\ast0}\bar{D}^{0}+h.c.}(s)+\Sigma_{D^{\ast+}%
D^{-}+h.c.}(s)\right]  +\lambda_{gg} \label{psi1omega}%
\end{equation}
with $s=m^{2}.$ Note, in the latter case the two-gluon contribution mentioned
previously has been formally included into the parameter  $\lambda_{gg}$.

The real part of the loops is depicted in Fig.~\ref{f5}. Equation (\ref{psi1rho}) is not
exactly zero because the loops $\Sigma_{D^{\ast0}\bar{D}^{0}+h.c.}(s)$ and
$\Sigma_{D^{\ast+}D^{-}+h.c.}(s)$ differ, due to the small mass
difference between the neutral and charged mesons $D$ and $D^{\ast}$ (in turn,
caused by the small mass difference between the quarks $u$ and $d$). Nevertheless,
when $s$ is far from the $D^{\ast0}D^{0}$ and $D^{\ast+}D^{-}$
thresholds, the difference between the loops in Eq.~(\ref{psi1rho}) is very small (see Fig.~\ref{f5}). At the mass of the state coming from the seed in case I, i.e.~$3.99$
GeV, the coupling to $\psi(1S)\rho$ (Eq.~\eqref{psi1rho}) can be safely neglected. On the contrary, as shown in Fig.~\ref{f5},
for the mass of the $X(3872)$ the situation is different: the
difference between the $D^{\ast0}\bar{D}^{0}$ and the
$D^{\ast+}D^{-}$ loops is non-negligible and quite important.

In what concerns the coupling $\xi_{\chi_{_{c1}}(2P)\rightarrow\psi(1S)\omega}$ in Eq.~(\ref{psi1omega}), the sum of the loops ensures that it is large for both the $X(3872)$ and the seed state. Here, as a first approximation, we
shall neglect the direct two-gluon contribution $\lambda_{gg}$, since close
to $DD^{\ast}$ thresholds the loops are large (the real parts have a peak in
that energy region, see Fig.~\ref{f5}), and the coupling constant $g_{\chi
_{c1}DD^{\ast}}$ is sizable (see also the recent argument in Ref.~\cite{milenapeter}). 
At $m=3.872$ GeV, corresponding to the $X(3872)$, one has the following ratio:
\begin{equation}
\left\vert \frac{\xi_{\chi_{_{c1}}(2P)\rightarrow\psi(1S)\omega}(m)}{\xi
_{\chi_{_{c1}}(2P)\rightarrow\psi(1S)\rho}(m)}\right\vert _{m=3.872\text{
GeV}}^{2}\simeq\left\vert \frac{\Sigma_{D^{\ast0}\bar{D}^{0}+h.c.}%
(s)+\Sigma_{D^{\ast+}D^{-}+h.c.}(s)}{\Sigma_{D^{\ast0}\bar{D}^{0}+h.c.}%
(s)-\Sigma_{D^{\ast+}D^{-}+h.c.}(s)}\right\vert _{s=3.872\text{ GeV}^{2}}%
^{2}=12.3\text{ .}%
\end{equation}
Thus, the coupling $\xi_{\chi_{_{c1}}(2P)\rightarrow\psi(1S)\rho}$ is indeed non-negligible (even if, as
expected, suppressed). On the other hand, for $m=3.986$ GeV (at the broad seed peak) one finds a ratio of about $630$, thus showing that the decay to
$\psi(1S)\rho$ is heavily suppressed (the loops cancel to a very good extent).
Thus, for the seed state at about $3.986$ GeV there is only the decay into
$\psi(1S)\omega.$

The decay widths of $\chi_{_{c1}}(2P)$, with a running mass $m$, into $\omega$
(or $\rho$), with a running mass $x$, can be summarized as:%

\begin{align}
\Gamma_{\chi_{_{c1}}(2P)\rightarrow\psi(1S)\omega}(m,x)  &  =\xi_{\chi_{_{c1}%
}(2P)\rightarrow\psi(1S)\omega}^{2}V(m,x)\text{ ,}\\
\Gamma_{\chi_{_{c1}}(2P)\rightarrow\psi(1S)\rho}(m,x)  &  =\xi_{\chi_{_{c1}%
}(2P)\rightarrow\psi(1S)\rho}^{2}V(m,x)\text{ ,}%
\end{align}
with%

\begin{equation}
V(m,x)=\frac{k}{8\pi m^{2}}\frac{1}{3m_{\psi(1S)}^{2}}\left[  4k^{4}%
+6m_{\psi(1S)}^{2}x^{2}+2k^{2}\left(  2m_{\psi(1S)}^{2}+x^{2}+2\sqrt
{k^{2}+m_{\psi(1S)}^{2}}\sqrt{k^{2}+x^{2}}\right)  \right]  \text{ ,}%
\end{equation}
where $k=k(m,m_{\psi(1S)},x).$ Then, by fixing the mass to $m=3.872$ GeV, and
further integrating over the $\rho$ mass, the decay $X(3872)\rightarrow
\psi(1S)\rho^{0}\rightarrow\psi(1S)\pi^{+}\pi^{-}$reads:%
\begin{equation}
\Gamma_{X(3872)\rightarrow\psi(1S)\pi^{+}\pi^{-}}=\left\vert \xi_{\chi_{_{c1}%
}(2P)\rightarrow\psi(1S)\rho}(m_{X(3872)})\right\vert ^{2}\int_{0}^{\infty
}\mathrm{d}x\ V(m_{X(3872)},x)d_{\rho}(x)\text{ },
\end{equation}
where $d_{\rho}(x)$ is the spectral function of the $\rho$ meson. Here, we
shall use a relativistic Breit-Wigner function:%
\begin{equation}
d_{\rho^{0}}(x)=N_{\rho}\frac{\theta(x-2m_{\pi^{+}})}{(x^{2}-m_{\rho^{0}}%
^{2})^{2}+\Gamma_{\rho^{0}}^{2}m_{\rho^{0}}^{2}}\ , \label{rhosf}%
\end{equation}
with the parameters $m_{\rho^{0}}=775.26\pm0.25$ MeV, $\Gamma_{\rho^{0}}=147.8\pm0.9$ MeV, and
$N_{\rho}$ such that $\int_{0}^{\infty}\mathrm{dx}\ d_{\rho}(x)=1.$

Similarly, the decay $X(3872)\rightarrow\psi(1S)\omega\rightarrow\psi
(1S)\pi^{+}\pi^{-}\pi^{0}$ is obtained upon integrating over the $\omega$
mass:
\begin{equation}
\Gamma_{X(3872)\rightarrow\psi(1S)\pi^{+}\pi^{-}\pi^{0}}=\left\vert \xi
_{\chi_{_{c1}}(2P)\rightarrow\psi(1S)\omega}(m_{X(3872)})\right\vert ^{2}%
\int_{0}^{\infty}\mathrm{d}x\ V(m_{X(3872)},x)d_{\omega}(x)\ ,
\end{equation}
where
\begin{equation}
d_{\omega}(x)=N_{\omega}\frac{\theta(x-2m_{\pi^{+}}-m_{\pi^{0}})}%
{(x^{2}-m_{\omega}^{2})^{2}+\Gamma_{\omega}^{2}m_{\omega}^{2}}\ ,
\end{equation}
with $m_{\omega}=782.65\pm0.12$ MeV, $\Gamma_{\omega}=8.49\pm0.08$ MeV, and
$N_{\omega}$ such that $\int_{0}^{\infty}\mathrm{dx}\ d_{\omega}(x)=1.$

Finally, the following ratio is obtained:
\begin{equation}
\frac{\Gamma_{X(3872)\rightarrow\psi(1S)\pi^{+}\pi^{-}\pi^{0}}}{\Gamma
_{X(3872)\rightarrow\psi(1S)\pi^{+}\pi^{-}}}\simeq1.9 \label{ratio}%
\end{equation}
which is of the same order of the experimental value $0.8\pm0.3$ listed in the \cite{pdg}. Summarizing, the originally small isospin breaking coupling to $J/\psi\rho$
is enhanced at the mass of the $X(3872)$, due to the difference between the real parts of
the neutral and charged loops. In this way, a qualitative agreement with data
is obtained. This is a surprisingly good result obtained without any further
assumptions and free parameters. Indeed, aslo in Refs. \cite{ferretti,xiao2,gamermann2} the small isopsin breaking is enhanced by analogous loop phenomena. 

Nevertheless, it should be stressed that our result is close
but not yet in full agreement with the experimental data. Within this context,
future improvements are possible: the phenomenological Lagrangian of
Eq.~(\ref{DDjor}), that describes the transition $DD^{\ast}$ into
$J/\psi\omega$ and $J/\psi\rho,$ is a useful but still rather simple
approximation (small isospin breaking could be included already at this
level). Moreover, the small but nonzero role of the direct two-gluon process,
parametrized by $\lambda_{gg}$ in\ Eq.~(\ref{psi1omega}), can slightly change
the result: for instance, a slight destructive interference would decrease the
coupling into the $J/\psi\omega$ channel. A better spectral function of the
$\rho$-meson, which goes beyond the relativistic Breit-Wigner approximation of
Eq.~(\ref{rhosf}), would also lead to a smaller ratio due to the fact that a
greater weight appears at lower energies, once a realistic form factor for the
$\rho$ is taken into account. Yet, such a procedure would imply the need of
additional parameters and would only lead to small changes, therefore it is
not considered in this work. Finally, the ratio in Eq.~(\ref{ratio}) is
reported for different values in the Appendix \ref{ap}. In the future, one
should included all these effects for a more quantitative description of the
decay ratio of Eq. (47), but the main idea -the isospin breaking enhancement
due to loops close to threshold- will still be the dominant and most
interesting contribution.

In conclusion, some important $D^{0}D^{\ast0}$ molecular properties for the
state $X(3872)$ naturally appear, as it is expected for a state whose pole is
so close to the $D^{0}D^{\ast0}$ threshold. At the same time, within our
framework, the state $X(3872)$ is intimately connected to the charmonium state
$\chi_{c1}(2P)$ (intuitively speaking, both states and poles form a unique
object): the state $X(3872)$ would not emerge if the coupling of the
$\chi_{c1}(2P)$ to $DD^{\ast}$ were too small or if its mass were too large
(see also results for parameter variation in the Appendix).

\begin{figure}[h!]
\begin{center}
\includegraphics[width=0.8 \textwidth]{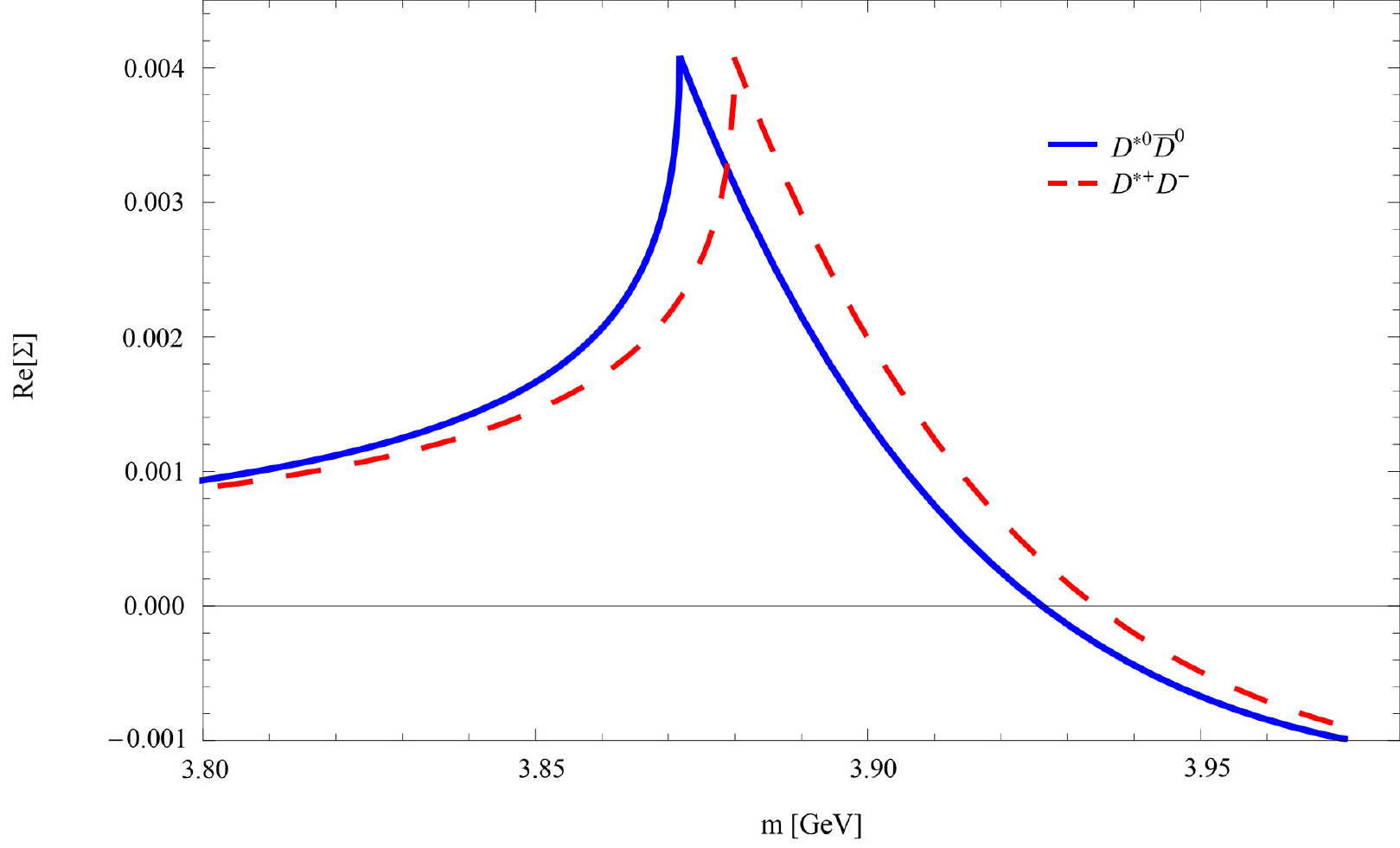}
\caption{\label{f5}Real part of the loop functions $\Sigma_{D^{0}D^{\ast0}}(m^{2})$ and
$\Sigma_{D^{+}D^{\ast-}}(m^{2})$, see Eq. (\ref{sigmas}). The peaks at the
thresholds are a salient features of these objects. The difference between
them in the energy region close to the thresholds is visbile: this is in the
end responsbile for the isopsin suppressed decay into $J/\psi\rho.$}
\end{center}
\end{figure}


\bigskip

\bigskip

\section{\label{s4}Conclusions}

In this work, we have studied the charmonium state $\chi_{c1}(2P)$ by using a
(relativistic) QFT Lagrangian. The coupling of this state to $DD^{\ast}$
mesons is quite strong and generates an additional enhanced peak in the
spectral function close to the $D^{0}D^{\ast0}$ state: in agreement with
various previous works on the subject (e.g. Refs.
[\cite{kalashnikova,gutschedong,gamermann,stapleton,hanhart,kalashnikova2,ortega,coito3872,santopinto,ferretti,cardoso,takeuchi,xiao1,xiao2}%
] and refs. therein), we assign this peak to the famous state $X(3872),$ that
-as a consequence- shows some features compatible with a $\bar{c}c$ state,
such as radiative decays and prompt production, and other features compatible
with a $D^{0}D^{\ast0}$ molecular state, such as the mass close to the
$D^{0}D^{\ast0}$ threshold and an isospin-breaking decay into $J/\psi\rho.$ 

In the complex plane, within our framework we find two poles: a standard seed
pole in the III RS corresponding to the predominantly $c\bar{c}$ object
$\chi_{c1}(2P),$ with a total strong decay width of about $80$ MeV (in
agreement with the predictions of the quark model), and, in addition, a
virtual pole on the II RS just below $D^{0}D^{\ast0}$ threshold, responsible
for the high peak of the spectral function at threshold, identified with
$X(3872)$. This second pole is regarded as a dynamically generated `companion'
pole, in agreement with analogous studies in the low-energy sector. (Within
our framework, we have also made independent predictions for the strong decay
to $D^{0}D^{\ast0}$, for the radiative decays to $\psi(1S)\gamma$ and
$\psi(2S)\gamma$, and for the ratio between the decays to $J/\psi\rho$ and
$J/\psi\omega,$ that can be useful- together with results of other authors on
the subject- to interpret future experimental outcomes.

The overall spectral function of $\chi_{c1}(2P)$, correctly normalized to
unity, describes simultaneously the state coming from the seed, i.e., the
$\chi_{c1}(2P)$, and the $X(3872).$ The $\bar{c}c$ seed state corresponds to a
large bump on the right side of the spectral function, but for some parameter
choices the bump disappears (scenario 2), eventually explaining why this state
was not yet measured. In fact, if -as the most authors agree- the $X(3872)$ is
predominantly not a standard charmonium, we should understand which properties
the bare charmonium has in order to find it in future experiments or to
understand why it does not show up. In general, we regard our results for the
pole position (hence mass and width) and (some of) the properties of the
predominantly charmonium state ,$\chi_{c1}(2P)$ such as the partial decay
width, as one of the main outcomes of our work. 

As a future outlook of our work, one should consider additional decay channels
in the evaluation of the propagator of the state $\chi_{c1}(2P)$: even if the
$DD^{\ast}$ mode is dominant, other channels could be important to better
understand the shape close to the left threshold, thus could improve the
understanding of $X(3872).$

In conclusion, the state $X(3872)$ emerges from a very peculiar interplay
between a quark-antiquark axial-vector state, and the thresholds placed at a
critical distance. It is then not surprising that no analogous state has been
found in the bottomonium sector
\cite{nopartneratlas,nopartnercms,nopartnerbelle,nopartnerlhcb}. In fact, the
existence of the $X(3872)$, as part of the spectral function of the $\chi
_{c1}(2P)$, emerges from the fulfillment of specific conditions, most notably
the proper distance to the relevant thresholds and the corresponding
couplings. Going from the charm to the bottomonium sector is likely to ruin
these somewhat delicate conditions at least for the analogous system, since
the bottomonium states are typically more bound than the charmonium ones, and
the second and third radial excitations of the axial-vector bottomonium are
unlikely to be close enough to their respective $s$-wave $BB^{\ast}$ decay
channel. Yet, it is possible, see e.g. Refs. \cite{ferretti,xiao3} that other
states analogous to the $X(3872)$ emerge in other parts of the rich QCD
spectrum. 

\bigskip

\textbf{Acknowledgements} F.G.~thanks A.~Pilloni for useful discussions. The authors acknowledge financial support from the
Polish National Science Centre (NCN) through the OPUS project no.
2015/17/B/ST2/01625. F. G. thanks also financial support from the Polish National Science Centre (NCN) through the OPUS project no. 2018/29/B/ST2/02576.


\appendix

\section{\label{ap}Results for parameter variation}

In this appendix we check how the results of our model vary upon changing the
free parameters. In Tables \ref{I}-\ref{VIII}, we report the main outcomes of our
approach. For each case tested here we show the numerical values of the
following quantities: the coupling constant $g_{\chi_{c1}DD^{\ast}}%
$; the probability that the dressed state $\chi_{c1}(2P)$ is contained between the left threshold and $1$ MeV above it, see Eq.~(\ref{integralthr}); the decay width into $D^{\ast0}\bar{D}^{0}$, see Eq.~(\ref{intwidth}); the pole positions in the complex energy plane; the
radiative decays reported in Eqs.~(\ref{raddec1}) and (\ref{raddec2}); the
ratio of Eq.~(\ref{ratio}) involving the isospin-breaking decay into
$J/\psi\rho^{0}$.

\subsection{\label{ap1}Variation of $\mathbf{g_{\chi_{c1}DD^{*}}}$ at fixed
$\mathbf{\Lambda}$, tested for two types of form factors}

\label{appg} As a first step, for the fixed cutoff value $\Lambda=0.5$ GeV,
we test different values of the coupling constant $g_{\chi_{c1}DD^{\ast}},$
obtained by changing the mass $m_{\ast}$, defined in\ Eq.~(\ref{mstar}), in the
range between $D^{\ast0}D^{0}$ and $D^{\ast+}D^{-}$ thresholds. The
critical value of $g_{\chi_{c1}DD^{\ast}}$ corresponds to the case when
$m_{\ast}$ sits just at the $D^{\ast0}D^{0}$ threshold (thus, the
virtual pole has a mass just below $m_{D_{0}}+m_{D_{0}^{\ast}}$, if the coupling is 
infinitesimally smaller than the critical value). Moreover, we also study one
case in which $m_{\ast}$ lies below $D^{\ast0}D^{0}$: here, there is a
real bound state and the spectral function has the form given by Eq.~(\ref{realpoleds}). 

In what concerns the bare mass of the charmonium,
we use the two values employed for the cases I and II described in the text, viz.~$m_{0}=3.95$ GeV \cite{isgur} in Sec.~\ref{sectionggauss} and $m_{0}=3.92$
GeV \cite{ebert} in Sec.~\ref{sectiongdipolar}.

Moreover, we have repeated all the calculations for two types of form factors,
the first one given by the Gaussian function in Eq.~(\ref{gaussff}), while the
second one has the form:
\begin{equation}
F_{\Lambda}\equiv F_{\Lambda}^{\text{dipolar}}(k)=\left(  1+\frac{k^{4}%
}{\Lambda^{4}}\right)  ^{-2}\text{.}\label{dipolarformfactor}%
\end{equation}

The variation of the coupling constant $g_{\chi_{c1}DD^{\ast}}$, as well as the specific choice of  
the form factor, either \eqref{gaussff} or \eqref{dipolarformfactor}, does not affect the overall picture, which is quite stable and
consistent for the presented results. In each case, there is a strong
enhancement close to the $D^{\ast0}D^{0}$ threshold and there are two
poles in the complex plane, one corresponding to the seed state and one to the
$X(3872)$ (which is virtual when the coupling constant is smaller than the
critical value, and real otherwise). The pole corresponding to the $X(3872)$ (indicated by $\blacklozenge$ in the tables) is a virtual pole (that is, in the II RS) when $g_{\chi_{c1}DD^{\ast}}$ is smaller than a critical value, but a bound-state (on the I RS) when  $g_{\chi_{c1}DD^{\ast}}$ exceeds the critical value. This is the case for Tables \ref{I}-\ref{IV}.

It is interesting to note that the state $X(3872)$ fades away when the coupling constant decreases. This property confirms that the very existence of the $X(3872)$ is due to
the nearby quarkonium state. Interestingly, a similar conclusion was achieved in Ref.~\cite{coito3872} and \cite{cardoso}, where the $X(3872)$ always required a small charmonium component to exist. However, the predictions concerning the $\chi_{c1}(2P)$ diverge from the present approach. In these references, by employing potential models with a harmonic oscillator as a confinement potential, the state that is found between 4.0 and 4.1 GeV is much larger that the pole at 3.99 GeV within the present approach, so that it would hardly be observed in the experiment. On the other hand, either for a seed value of 3.95 GeV, or by the inclusion of additional decay channels, the $X(3872)$ becomes a seed pole, rather than a dynamically generated one. Such possibility should be disentangled by a possible experimental observation of the $\chi_{c1}(2P)$ state, as we present here.  Another important difference is that, within the present paper, we can understand radiative and isospin-violating decay of $X(3872)$ without additional assumptions, see Sec. \ref{s3}.

\newpage

\subsubsection{\label{sectionggauss} $\mathbf{m_{0}=3.95}$ \textbf{GeV}}
The results for $m_{0}=3.95$ GeV are presented in Table
\ref{I} (Gaussian form factor) and Table \ref{II} (dipolar form factor). 

\begin{table}[h!]
\renewcommand{\arraystretch}{1.20}\label{tableggauss}
\begin{tabular}[c]{|c|c|c|l|c|c|c|}
\hline
\multicolumn{7}{|c|}{\textbf{Gaussian form factor, \hspace{0.5cm} $\mathbf{\Lambda=0.5}$ GeV, \hspace{0.5cm} $\mathbf{m_0=3.95}$ \textbf{GeV,} \hspace{0.5cm} $\mathbf{m_* \neq}$ \textbf{const.}}}\\
\hline
$g_{\chi_{c1}DD^*}$ & Eq.~(\ref{integralthr})  &Eq.~(\ref{intwidth})& \hspace{0.7cm}Pole positions & Eq.~(\ref{raddec1}) & Eq.~(\ref{raddec2}) & Eq.~(\ref{ratio})\\

[GeV]& &[MeV]&\hspace{1.2cm} [GeV] & [keV]& [keV] & \\
\hline
\hline
9.808&0.057&0.636&$\bullet  \hspace{0.5cm} 3.9961-0.0359$ $i$&0.628&3.64&1.92\\
(critical)&&&$\blacklozenge \hspace{0.45cm} 3.8717-i\varepsilon$&&&\\
\hline
9.732&0.049&0.607&$\bullet  \hspace{0.5cm} 3.9954-0.0357$ $i$&0.539&3.13&1.92\\
(case I)&&&$\blacklozenge \hspace{0.45cm} 3.8716-i\varepsilon$&&&\\
\hline
9.500&0.029&0.408&$\bullet  \hspace{0.5cm} 3.9933-0.0354$ $i$&0.323&1.88&1.92\\
&&&$\blacklozenge \hspace{0.45cm} 3.8715-i\varepsilon$&&&\\
\hline
9.300&0.019&0.263&$\bullet  \hspace{0.5cm} 3.9915-0.0350$ $i$&0.206&1.20&1.92\\
&&&$\blacklozenge \hspace{0.45cm} 3.8710-i\varepsilon$&&&\\
\hline
9.000&0.010&0.136&$\bullet  \hspace{0.5cm} 3.9887-0.0344$ $i$&0.110&0.64&1.92\\
&&&$\blacklozenge \hspace{0.45cm} 3.8699-i\varepsilon$&&&\\
\hline
8.800&0.007&0.091&$\bullet  \hspace{0.5cm} 3.9869-0.0339$ $i$&0.076&0.44&1.92\\
&&&$\blacklozenge \hspace{0.45cm} 3.8689-i\varepsilon$&&&\\
\hline
8.000&0.002&0.024&$\bullet  \hspace{0.5cm} 3.9796-0.0316$ $i$&0.024&0.14&1.92\\
&&&$\blacklozenge \hspace{0.45cm} 3.8609-i\varepsilon$&&&\\
\hline
10.000&0.035&0.505&$\bullet  \hspace{0.5cm} 3.9978-0.0361$ $i$&0.387&2.25&1.92\\
&&&$\blacklozenge \hspace{0.45cm} 3.8716-i\varepsilon$ \hspace{0.1cm} (I RS)&&&\\ 
\hline
\end{tabular}
\caption{\label{I}Results for different values of the coupling constant $g_{\chi
_{c1}DD^{\ast}}$, calculated for $\Lambda=0.5$ GeV, bare mass $m_{0}=3.95$ GeV,
and Gaussian vertex function. The symbols $(\bullet)$ and $(\blacklozenge)$
indicate the seed pole and the virtual companion pole,
respectively.}
\end{table}
\vspace*{-0mm}
\begin{table}[h!]
\label{tablegdipolar}
\renewcommand{\arraystretch}{1.20}
\begin{tabular}[c]{|c|c|c|l|c|c|c|}
\hline
\multicolumn{7}{|c|}{\textbf{Dipolar form factor, \hspace{0.5cm} $\mathbf{\Lambda=0.5}$ GeV, \hspace{0.5cm} $\mathbf{m_0=3.95}$ \textbf{GeV,} \hspace{0.5cm} $\mathbf{m_* \neq}$ \textbf{const.}}}\\
\hline
$g_{\chi_{c1}DD^*}$  &Eq.~(\ref{integralthr})&Eq.~(\ref{intwidth})& \hspace{0.7cm}Pole positions & Eq.~(\ref{raddec1}) & Eq.~(\ref{raddec2})  & Eq.~(\ref{ratio})\\

[GeV]& &[MeV]&\hspace{1.2cm} [GeV] & [keV]& [keV]& \\
\hline
\hline
8.339&0.078&0.630&$\bullet  \hspace{0.5cm} 4.0075-0.0390$ $i$&0.856&4.97&2.87\\
(critical)&&&$\blacklozenge \hspace{0.45cm} 3.8717-i\varepsilon$&&&\\
\hline
8.179&0.047&0.481&$\bullet  \hspace{0.5cm} 4.006-0.0389$ $i$&0.520&3.02&2.87\\
&&&$\blacklozenge \hspace{0.45cm} 3.8715-i\varepsilon$&&&\\
\hline
7.800&0.013&0.138&$\bullet  \hspace{0.5cm} 4.001-0.0385$ $i$&0.146&0.85&2.87\\
&&&$\blacklozenge \hspace{0.45cm} 3.8675-i\varepsilon$&&&\\
\hline
7.500&0.060&0.059&$\bullet  \hspace{0.5cm} 3.9980-0.0381$ $i$&0.066&0.38&2.87\\
&&&$\blacklozenge \hspace{0.45cm} 3.8616-i\varepsilon$&&&\\
\hline
7.200&0.0032&0.029&$\bullet  \hspace{0.5cm} 3.9945-0.0376$ $i$&0.35&0.21&2.87\\
&&&$\blacklozenge \hspace{0.45cm} 3.8628-i\varepsilon$&&&\\
\hline
7.000&0.0022&0.020&$\bullet  \hspace{0.5cm} 3.9921-0.0373$ $i$&0.025&0.15&2.87\\
&&&$\blacklozenge \hspace{0.45cm} 3.8707-i\varepsilon$&&&\\
\hline
6.500&0.0011&0.008&$\bullet  \hspace{0.5cm} 3.9861-0.0361$ $i$&0.012&0.072&2.87\\
&&&$\blacklozenge \hspace{0.45cm} 3.8658-i\varepsilon$&&&\\
\hline
8.500&0.039&0.417&$\bullet  \hspace{0.5cm} 4.009-0.0391$ $i$&0.426&2.48&2.87\\
&&&$\blacklozenge \hspace{0.45cm} 3.8716-i\varepsilon$ \hspace{0.1cm} (I RS)&&&\\
\hline
\end{tabular}
\caption{\label{II}Similar to Table \ref{I}, but for dipolar form factor.}%
\end{table}

\subsubsection{\label{sectiongdipolar} $\mathbf{m_{0}=3.92}$ \textbf{GeV}}

The results for $m_{0}=3.92$ GeV are presented in
Table \ref{III} (Gaussian form factor) and Table \ref{IV} (dipolar form factor).

\begin{table}[h!]
\renewcommand{\arraystretch}{1.20}
\begin{tabular}[c]{|c|c|c|l|c|c|c|}
\hline
\multicolumn{7}{|c|}{\textbf{Gaussian form factor, \hspace{0.5cm} $\mathbf{\Lambda=0.5}$ GeV, \hspace{0.5cm} $\mathbf{m_0=3.92}$ \textbf{GeV,} \hspace{0.5cm} $\mathbf{m_* \neq}$ \textbf{const.}}}\\
\hline
$g_{\chi_{c1}DD^*}$ &Eq.~(\ref{integralthr}) &Eq.~(\ref{intwidth})& \hspace{0.7cm}Pole positions & Eq.~(\ref{raddec1}) & Eq.~(\ref{raddec2}) & Eq.~(\ref{ratio})\\

[GeV]&&[MeV]&\hspace{1.2cm} [GeV] & [keV]& [keV] & \\
\hline
\hline
7.689&0.092&0.634&$\bullet  \hspace{0.5cm} 3.9547-0.0444$ $i$&1.02&5.91&1.92\\
(critical)&&&$\blacklozenge \hspace{0.45cm} 3.8717-i\varepsilon$&&&\\
\hline
7.557&0.067&0.544&$\bullet  \hspace{0.5cm} 3.9531-0.0440$ $i$&0.74&4.27&1.92\\
(case II)&&&$\blacklozenge \hspace{0.45cm} 3.8716-i\varepsilon$&&&\\
\hline
7.400&0.043&0.373&$\bullet  \hspace{0.5cm} 3.9513-0.0435$ $i$&0.48&2.77&1.92\\
&&&$\blacklozenge \hspace{0.45cm} 3.8713-i\varepsilon$&&&\\
\hline
7.300&0.033&0.283&$\bullet  \hspace{0.5cm} 3.9501-0.0432$ $i$&0.36&2.08&1.92\\
&&&$\blacklozenge \hspace{0.45cm} 3.8710-i\varepsilon$&&&\\
\hline
7.000&0.015&0.123&$\bullet  \hspace{0.5cm} 3.9465-0.0420$ $i$&0.16&0.95&1.92\\
&&&$\blacklozenge \hspace{0.45cm} 3.8689-i\varepsilon$&&&\\
\hline
6.800&0.009&0.075&$\bullet  \hspace{0.5cm} 3.9441-0.0412$ $i$&0.10&0.60&1.92\\
&&&$\blacklozenge \hspace{0.45cm} 3.8664-i\varepsilon$&&&\\
\hline
6.600&0.006&0.048&$\bullet  \hspace{0.5cm} 3.9417-0.0402$ $i$&0.07&0.41&1.92\\
&&&$\blacklozenge \hspace{0.45cm} 3.8629-i\varepsilon$&&&\\
\hline
8.000&0.034&0.340&$\bullet  \hspace{0.5cm} 3.9582-0.0452$ $i$&0.38&2.19&1.92\\
&&&$\blacklozenge \hspace{0.45cm} 3.8714-i\varepsilon$ \hspace{0.1cm} (I RS)&&&\\
\hline
\end{tabular}
\caption{\label{III}Results for different values of the coupling constant $g_{\chi
_{c1}DD^{\ast}}$, calculated for $\Lambda=0.5$ GeV, bare mass $m_{0}=3.92$ GeV,
and Gaussian vertex function. The symbols $(\bullet)$ and $(\blacklozenge)$
indicate the seed pole and the virtual companion pole,
respectively.}%
\end{table}

\begin{table}[h!]
\renewcommand{\arraystretch}{1.20}
\begin{tabular}[c]{|c|c|c|l|c|c|c|}
\hline
\multicolumn{7}{|c|}{\textbf{Dipolar form factor, \hspace{0.5cm} $\mathbf{\Lambda=0.5}$ GeV, \hspace{0.5cm} $\mathbf{m_0=3.92}$ \textbf{GeV,} \hspace{0.5cm} $\mathbf{m_* \neq}$ \textbf{const.}}}\\
\hline
$g_{\chi_{c1}DD^*}$ & Eq.~(\ref{integralthr})&Eq.~(\ref{intwidth})& \hspace{0.7cm}Pole position & Eq.~(\ref{raddec1}) & Eq.~(\ref{raddec2})  & Eq.~(\ref{ratio})\\

[GeV]&&[MeV]&\hspace{1.2cm} [GeV] & [keV]& [keV] & \\
\hline
\hline
6.537&0.125&0.62&$\bullet  \hspace{0.5cm} 3.9700-0.0498$ $i$&1.38&8.00&2.87\\
(critical)&&&$\blacklozenge \hspace{0.45cm} 3.8717-i\varepsilon$&&&\\
\hline
6.351&0.062&0.40&$\bullet  \hspace{0.5cm} 3.9674-0.0498$ $i$&0.68&3.94&2.87\\
&&&$\blacklozenge \hspace{0.45cm} 3.8712-i\varepsilon$&&&\\
\hline
6.000&0.015&0.094&$\bullet  \hspace{0.5cm} 3.9624-0.0498$ $i$&0.16&0.96&2.87\\
&&&$\blacklozenge \hspace{0.45cm} 3.8678-i\varepsilon$&&&\\
\hline
5.800&0.081&0.048&$\bullet  \hspace{0.5cm} 3.95934-0.0497$ $i$&0.09&0.52&2.87\\
&&&$\blacklozenge \hspace{0.45cm} 3.8682-i\varepsilon$&&&\\
\hline
5.500&0.0039&0.021&$\bullet  \hspace{0.5cm} 3.9546-0.0495$ $i$&0.043&0.25&2.87\\
&&&$\blacklozenge \hspace{0.45cm} 3.8688-i\varepsilon$&&&\\
\hline
5.200&0.0022&0.011&$\bullet  \hspace{0.5cm} 3.9495-0.0492$ $i$&0.024&0.14&2.87\\
&&&$\blacklozenge \hspace{0.45cm} 3.8694-i\varepsilon$&&&\\
\hline
4.500&0.0008&0.0029&$\bullet  \hspace{0.5cm} 3.9355-0.0477$ $i$&0.0088&0.051&2.87\\
&&&$\blacklozenge \hspace{0.45cm} 3.8960-i\varepsilon$&&&\\
\hline
6.800&0.032&0.23&$\bullet  \hspace{0.5cm} 3.9735-0.0498$ $i$&0.35&2.03&2.87\\
&&&$\blacklozenge \hspace{0.45cm} 3.8713-i\varepsilon$ \hspace{0.1cm} (I RS)&&&\\
\hline
\end{tabular}
\caption{\label{IV}Similar to Table \ref{III} but for a dipolar form factor.}%
\end{table}

\subsection{\label{ap2}Variation of $\mathbf{\Lambda}$ at fixed $\mathbf{m_\ast}$, tested
for two types of form factors}

For completeness, we study the dependence of the results on the parameter
$\Lambda$. We test different $\Lambda$ in the range from $0.4$ GeV to $0.8$
GeV. Similarly to Sec.~\ref{ap1}, we use the two bare masses $m_{0}=3.95$ GeV
(Sec.~\ref{lambdaappa}) and $m_{0}=3.92$ GeV (Sec.~\ref{lambdaappb}), as well
as the Gaussian and dipolar form factors. In order to determine the coupling
constant $g_{\chi_{c1}DD^{\ast}},$ we set $m_{\ast}=3.874$ GeV (in between the two $DD^{\ast}$ thresholds, just as done in the main text).
One should notice that, even if $m_{\ast}$ is fixed, the coupling constant
varies. 

We conclude that the value of the cutoff in the quite large range from
$0.4$ GeV to $0.8$ GeV does not change the most important features emerging from our
approach. For each value of $\Lambda$, two poles are observed on the complex
energy plane. However, the value $\Lambda=0.8$ GeV should be regarded as an upper limit, since
the imaginary part of the pole corresponding to the standard seed state gives
rise to a too large decay width when compared with the predictions of the quark
model with a Cornell potential. For $m_{0}=3.95$ GeV, the obtained decay widths are $\sim225$ MeV and
$\sim292$ MeV for Gaussian and dipolar form factors, respectively. Similarly,
for $m_{0}=3.92$ GeV we get $\sim217$ MeV and $\sim359$ MeV for these two form
factors. In such cases, the predominantly $\bar{c}c$ would be very difficult to detect in the experiment, but the width would be similar to the one obtained in quark model approaches which employ an harmonic oscillator potential \cite{coito3872}.

\newpage

\subsubsection{\label{lambdaappa}$\mathbf{m_{0}=3.95}$ \textbf{GeV}}

The results for $m_{0}=3.95$ GeV are presented in Table \ref{V} (Gaussian form factor) and Table \ref{VI} (dipolar form factor). 

\begin{table}[h!]
\renewcommand{\arraystretch}{1.20}
\begin{tabular}[c]{|c|c|c|c|l|c|c|c|}
\hline
\multicolumn{8}{|c|}{\textbf{Gaussian form factor, \hspace{0.5cm} $\mathbf{\Lambda\neq}$ \textbf{const.}, \hspace{0.5cm} $\mathbf{m_0=3.95}$ \textbf{GeV,} \hspace{0.5cm} $\mathbf{m_*=3.874}$ \textbf{GeV.} }}\\
\hline
$\Lambda$&$g_{\chi_{c1}DD^*}$ & Eq.~(\ref{integralthr}) &Eq.~(\ref{intwidth})& \hspace{0.7cm}Pole positions & Eq.~(\ref{raddec1})  & Eq.~(\ref{raddec2}) & Eq.~(\ref{ratio})\\

[GeV]&[GeV]&&[MeV]&\hspace{1.2cm} [GeV] & [keV]& [keV] & \\
\hline
\hline
0.4&11.259&0.040&0.629&$\bullet  \hspace{0.5cm} 3.9861-0.0171$ $i$&0.444&2.57&1.32\\
&&&&$\blacklozenge \hspace{0.45cm} 3.8717-i\varepsilon$&&&\\
\hline
0.42&10.897&0.045&0.636&$\bullet  \hspace{0.5cm} 3.9883-0.0204$ $i$&0.499&2.89&1.43\\
&&&&$\blacklozenge \hspace{0.45cm} 3.8717-i\varepsilon$&&&\\
\hline
0.45&10.413&0.048&0.632&$\bullet  \hspace{0.5cm} 3.9913-0.0258$ $i$&0.528&3.07&1.61\\
&&&&$\blacklozenge \hspace{0.45cm} 3.8717-i\varepsilon$&&&\\
\hline
0.5&9.732&0.049&0.607&$\bullet  \hspace{0.5cm} 3.9954-0.0357$ $i$&0.539&3.13&1.92\\
&(case I)&&&$\blacklozenge \hspace{0.45cm} 3.8716-i\varepsilon$&&&\\
\hline
0.55&9.169&0.050&0.577&$\bullet  \hspace{0.5cm} 3.9983-0.0468$ $i$&0.551&3.20&2.27\\
&&&&$\blacklozenge \hspace{0.45cm} 3.8716-i\varepsilon$&&&\\
\hline
0.6&8.694&0.051&0.549&$\bullet  \hspace{0.5cm} 3.9998-0.0588$ $i$&0.562&3.26&2.65\\
&&&&$\blacklozenge \hspace{0.45cm} 3.8716-i\varepsilon$&&&\\
\hline
0.7&7.930&0.053&0.497&$\bullet  \hspace{0.5cm} 3.9983-0.0848$ $i$&0.582&3.38&3.49\\
&&&&$\blacklozenge \hspace{0.45cm} 3.8715-i\varepsilon$&&&\\
\hline
0.8&7.338&0.054&0.454&$\bullet  \hspace{0.5cm} 3.9899-0.1123$ $i$&0.600&3.49&4.45\\
&&&&$\blacklozenge \hspace{0.45cm} 3.8715-i\varepsilon$&&&\\
\hline
\end{tabular}
\caption{\label{V}Results for different values of cutoff $\Lambda$. The used
parameters are: bare mass $m_{0}=3.95$ GeV, $m_{\ast}=3.874$ GeV and a Gaussian
vertex function. The symbols $(\bullet)$ and $(\blacklozenge)$ indicate the seed pole and the virtual companion pole, respectively.}%
\end{table}

\begin{table}[h!]
\renewcommand{\arraystretch}{1.20}
\begin{tabular}[c]{|c|c|c|c|l|c|c|c|}
\hline
\multicolumn{8}{|c|}{\textbf{Dipolar form factor, \hspace{0.5cm} $\mathbf{\Lambda\neq}$ \textbf{const.}, \hspace{0.5cm} $\mathbf{m_0=3.95}$ \textbf{GeV,} \hspace{0.5cm} $\mathbf{m_*=3.874}$ \textbf{GeV.} }}\\
\hline
$\Lambda$&$g_{\chi_{c1}DD^*}$  &Eq.~(\ref{integralthr})&Eq.~(\ref{intwidth})& \hspace{0.7cm}Pole positions & Eq.~(\ref{raddec1})& Eq.~(\ref{raddec2}) & Eq.~(\ref{ratio})\\

[GeV]&[GeV]&&[MeV]&\hspace{1.2cm} [GeV] & [keV]& [keV] & \\
\hline
\hline
0.4&9.369&0.0440&0.552&$\bullet  \hspace{0.5cm} 3.9909-0.0192$ $i$&0.485&2.82&1.90\\
&&&&$\blacklozenge \hspace{0.45cm} 3.8716-i\varepsilon$&&&\\
\hline
0.42&9.090&0.0447&0.536&$\bullet  \hspace{0.5cm} 3.9939-0.0226$ $i$&0.493&2.86&2.08\\
&&&&$\blacklozenge \hspace{0.45cm} 3.8716-i\varepsilon$&&&\\
\hline
0.45&8.714&0.0457&0.514&$\bullet  \hspace{0.5cm} 3.9984-0.0282$ $i$&0.503&2.92&2.36\\
&&&&$\blacklozenge \hspace{0.45cm} 3.8716-i\varepsilon$&&&\\
\hline
0.5&8.179&0.0472&0.481&$\bullet  \hspace{0.5cm} 4.0057-0.0389$ $i$&0.520&3.02&2.87\\
&&&&$\blacklozenge \hspace{0.45cm} 3.8715-i\varepsilon$&&&\\
\hline
0.55&7.732&0.0486&0.452&$\bullet  \hspace{0.5cm} 4.0126-0.0515$ $i$&0.536&3.12&3.43\\
&&&&$\blacklozenge \hspace{0.45cm} 3.8714-i\varepsilon$&&&\\
\hline
0.6&7.351&0.0499&0.427&$\bullet  \hspace{0.5cm} 4.0190-0.0660$ $i$&0.550&3.20&4.04\\
&&&&$\blacklozenge \hspace{0.45cm} 3.8714-i\varepsilon$&&&\\
\hline
0.7&6.732&0.0522&0.384&$\bullet  \hspace{0.5cm} 4.0298-0.1013$ $i$&0.575&3.34&5.42\\
&&&&$\blacklozenge \hspace{0.45cm} 3.8711-i\varepsilon$&&&\\
\hline
0.8&6.249&0.0539&0.348&$\bullet  \hspace{0.5cm} 4.0376-0.1458$ $i$&0.595&3.46&7.00\\
&&&&$\blacklozenge \hspace{0.45cm} 3.8706-i\varepsilon$&&&\\
\hline
\end{tabular}
\caption{\label{VI}Similar to Table \ref{V} but for a dipolar vertex function.}%
\end{table}

\subsubsection{$\mathbf{m_{0}=3.92}$ \textbf{GeV}}
\label{lambdaappb}
The results for $m_{0}=3.92$ GeV are presented in Tables \ref{VII} (Gaussian form factor), and \ref{VIII} (dipolar form factor).

\begin{table}[h!]
\renewcommand{\arraystretch}{1.20}
\begin{tabular}[c]{|c|c|c|c|l|c|c|c|}
\hline
\multicolumn{8}{|c|}{\textbf{Gaussian form factor, \hspace{0.5cm} $\mathbf{\Lambda\neq}$ \textbf{const.}, \hspace{0.5cm} $\mathbf{m_0=3.92}$ \textbf{GeV,} \hspace{0.5cm} $\mathbf{m_*=3.874}$ \textbf{GeV.} }}\\
\hline
$\Lambda$&$g_{\chi_{c1}DD^*}$ & Eq.~(\ref{integralthr})&Eq.~(\ref{intwidth})& \hspace{0.7cm}Pole positions & Eq.~(\ref{raddec1}) & Eq.~(\ref{raddec2}) & Eq.~(\ref{ratio})\\

[GeV]&[GeV]&&[MeV]&\hspace{1.2cm} [GeV] & [keV]& [keV] & \\
\hline
\hline
0.4&8.743&0.0662&0.626&$\bullet  \hspace{0.5cm} 3.9507-0.0246$ $i$&0.729&4.23&1.32\\
&&&&$\blacklozenge \hspace{0.45cm} 3.8717-i\varepsilon$&&&\\
\hline
0.42&8.461&0.0663&0.612&$\bullet  \hspace{0.5cm} 3.9517-0.0282$ $i$&0.730&4.24&1.43\\
&&&&$\blacklozenge \hspace{0.45cm} 3.8717-i\varepsilon$&&&\\
\hline
0.45&8.086&0.0664&0.587&$\bullet  \hspace{0.5cm} 3.9527-0.0339$ $i$&0.732&4.25&1.61\\
&&&&$\blacklozenge \hspace{0.45cm} 3.8716-i\varepsilon$&&&\\
\hline
0.5&7.557&0.0667&0.544&$\bullet  \hspace{0.5cm} 3.9531-0.0440$ $i$&0.735&4.27&1.92\\
&(case II)&&&$\blacklozenge \hspace{0.45cm} 3.8716-i\varepsilon$&&&\\
\hline
0.55&7.120&0.0670&0.504&$\bullet  \hspace{0.5cm} 3.9519-0.0548$ $i$&0.738&4.29&2.27\\
&&&&$\blacklozenge \hspace{0.45cm} 3.8716-i\varepsilon$&&&\\
\hline
0.6&6.751&0.0672&0.468&$\bullet  \hspace{0.5cm} 3.9490-0.0659$ $i$&0.741&4.31&2.65\\
&&&&$\blacklozenge \hspace{0.45cm} 3.8715-i\varepsilon$&&&\\
\hline
0.7&6.157&0.0675&0.408&$\bullet  \hspace{0.5cm} 3.9373-0.0883$ $i$&0.744&4.33&3.49\\
&&&&$\blacklozenge \hspace{0.45cm} 3.8713-i\varepsilon$&&&\\
\hline
0.8&5.698&0.0674&0.359&$\bullet  \hspace{0.5cm} 3.9172-0.1087$ $i$&0.744&4.33&4.45\\
&&&&$\blacklozenge \hspace{0.45cm} 3.8708-i\varepsilon$&&&\\
\hline
\end{tabular}
\caption{\label{VII}Results for different values of cutoff $\Lambda$. The used
parameters are: bare mass $m_{0}=3.92$ GeV, $m_{*}=3.874$ GeV, and a Gaussian
vertex function. The symbols $(\bullet)$ and $(\blacklozenge)$ indicate the seed
pole and the virtual companion pole, respectively.}%
\end{table}

\begin{table}[h!]
\renewcommand{\arraystretch}{1.20}
\begin{tabular}[c]{|c|c|c|c|l|c|c|c|}
\hline
\multicolumn{8}{|c|}{\textbf{Dipolar form factor, \hspace{0.5cm} $\mathbf{\Lambda\neq}$ \textbf{const.}, \hspace{0.5cm} $\mathbf{m_0=3.92}$ \textbf{GeV,} \hspace{0.5cm} $\mathbf{m_*=3.874}$ \textbf{GeV.} }}\\
\hline
$\Lambda$&$g_{\chi_{c1}DD^*}$ &Eq.~(\ref{integralthr})&Eq.~(\ref{intwidth})& \hspace{0.7cm}Pole positions & Eq.~(\ref{raddec1})& Eq.~(\ref{raddec2})  & Eq.~(\ref{ratio})\\

[GeV]&[GeV]&&[MeV]&\hspace{1.2cm} [GeV] & [keV]& [keV] & \\
\hline
\hline
0.4&7.275&0.060&0.483&$\bullet  \hspace{0.5cm} 3.9572-0.0265$ $i$&0.660&3.84&1.90\\
&&&&$\blacklozenge \hspace{0.45cm} 3.8715-i\varepsilon$&&&\\
\hline
0.42&7.058&0.060&0.463&$\bullet  \hspace{0.5cm} 3.9594-0.0305$ $i$&0.664&3.86&2.08\\
&&&&$\blacklozenge \hspace{0.45cm} 3.8715-i\varepsilon$&&&\\
\hline
0.45&6.766&0.061&0.437&$\bullet  \hspace{0.5cm} 3.9626-0.0371$ $i$&0.670&3.90&2.36\\
&&&&$\blacklozenge \hspace{0.45cm} 3.8714-i\varepsilon$&&&\\
\hline
0.5&6.351&0.062&0.399&$\bullet  \hspace{0.5cm} 3.9674-0.0498$ $i$&0.678&3.94&2.87\\
&&&&$\blacklozenge \hspace{0.45cm} 3.8712-i\varepsilon$&&&\\
\hline
0.55&6.004&0.062&0.366&$\bullet  \hspace{0.5cm} 3.9715-0.0647$ $i$&0.684&3.98&3.43\\
&&&&$\blacklozenge \hspace{0.45cm} 3.8709-i\varepsilon$&&&\\
\hline
0.6&5.708&0.062&0.338&$\bullet  \hspace{0.5cm} 3.9748-0.0819$ $i$&0.688&4.00&4.04\\
&&&&$\blacklozenge \hspace{0.45cm} 3.8703-i\varepsilon$&&&\\
\hline
0.7&5.228&0.0632&0.291&$\bullet  \hspace{0.5cm} 3.9790-0.1244$ $i$&0.692&4.03&5.42\\
&&&&$\blacklozenge \hspace{0.45cm} 3.8698-i\varepsilon$&&&\\
\hline
0.8&4.852&0.063&0.254&$\bullet  \hspace{0.5cm} 3.9813-0.1793$ $i$&0.690&4.01&7.00\\
&&&&$\blacklozenge \hspace{0.45cm} 3.8704-i\varepsilon$&&&\\
\hline
\end{tabular}
\caption{\label{VIII}Similar to Table \ref{VII} but for a dipolar vertex function.}%
\end{table}

\end{document}